\documentclass{ieeetj}
\newcommand{\seclogo}{}
\usepackage{cite}
\usepackage{amsmath,amssymb,amsfonts}
\usepackage{algorithmic}
\usepackage{graphicx,color}
\usepackage{textcomp}
\usepackage{xcolor}
\usepackage{hyperref}
\hypersetup{hidelinks=true}
\usepackage{algorithm,algorithmic}
\usepackage{subcaption}
\usepackage{booktabs}
\usepackage{soul}
\usepackage{caption}
\usepackage{multirow}
\usepackage{graphicx}
\usepackage{hyperref}
\usepackage{listings}
\let\labelindent\relax
\usepackage{enumitem}
\usepackage{placeins}
\usepackage{todonotes}
\usepackage{amssymb}
\usepackage{threeparttable}
\usepackage{diagbox}

\usepackage{graphicx}
\usepackage{adjustbox}

\def\BibTeX{{\rm B\kern-.05em{\sc i\kern-.025em b}\kern-.08em
    T\kern-.1667em\lower.7ex\hbox{E}\kern-.125emX}}
\AtBeginDocument{\definecolor{tmlcncolor}{cmyk}{0.93,0.59,0.15,0.02}\definecolor{NavyBlue}{RGB}{0,86,125}}

\def\authorrefmark#1{\ensuremath{^{\textbf{#1}}}}

\begin{document}
\receiveddate{XX Month, XXXX}
\reviseddate{XX Month, XXXX}
\accepteddate{XX Month, XXXX}
\publisheddate{XX Month, XXXX}
\currentdate{XX Month, XXXX}
\doiinfo{XXXX.2022.1234567}

\title{\huge MARVEL: An End-to-End Framework for Generating Model-Class Aware Custom RISC-V Extensions for Lightweight AI}

\author{Ajay Kumar M\authorrefmark{1}, Cian O'Mahoney\authorrefmark{1}, Pedro Kreutz Werle\authorrefmark{1}, Shreejith Shanker\authorrefmark{2}, Member, IEEE, Dimitrios S. Nikolopoulos\authorrefmark{3}, Fellow, IEEE, Bo Ji\authorrefmark{3}, Senior Member, IEEE, Hans Vandierendonck\authorrefmark{4}, Senior Member, IEEE, and Deepu John\authorrefmark{1}, Senior Member, IEEE}
\affil{University College Dublin, Dublin 4, Ireland}
\affil{Trinity College Dublin, Dublin 4, Ireland}
\affil{Virginia Tech, Blacksburg, VA 24061, USA}
\affil{Queen's University Belfast, Belfast, UK }
\corresp{Corresponding author: Ajay Kumar M (email: ajay.kumarm@ucdconnect.ie).}
\authornote{This work was funded by Taighde Éireann – Research Ireland  through the 1) Centre for Research Training in Machine Learning (18/CRT/6183), and 2) US-Ireland R\&D Partnership Programme (22/US/3848) and by MicroElectronic Circuit Centre Ireland}

\begin{abstract}
Deploying deep neural networks (DNNs) on resource-constrained IoT devices remains a challenging problem, often requiring hardware modifications tailored to individual AI models. Existing accelerator-generation tools, such as AMD's FINN, do not adequately address extreme resource limitations faced by IoT endpoints operating in bare-metal environments without an operating system (OS). To overcome these constraints, we propose MARVEL—an automated, end-to-end framework that generates custom RISC-V ISA extensions tailored to specific DNN model classes, with a primary focus on convolutional neural networks (CNNs). The proposed method profiles high-level DNN representations in Python and generates an ISA-extended RISC-V core with associated compiler tools for efficient deployment. The flow leverages (1) Apache TVM for translating high-level Python-based DNN models into optimized C code, (2) Synopsys ASIP Designer for identifying compute-intensive kernels, modeling, and generating a custom RISC-V and (3) Xilinx Vivado for FPGA implementation. Beyond a model-class specific RISC-V, our approach produces an optimized bare-metal C implementation, eliminating the need for an OS or extensive software dependencies. Unlike conventional deployment pipelines relying on TensorFlow/PyTorch runtimes, our solution enables seamless execution in highly resource-constrained environments. We evaluated the flow on popular DNN models such as LeNet-5*, MobileNetV1, ResNet50, VGG16, MobileNetV2 and DenseNet121 using the Synopsys trv32p3 RISC-V core as a baseline. Results show a \textbf{2$\times$} speedup in inference and upto \textbf{2$\times$} reduction in energy per inference at a 28.23\% area overhead when implemented on an AMD Zynq UltraScale+ ZCU104 FPGA platform. 
\end{abstract}


\maketitle
\section{INTRODUCTION}
The rapid proliferation of IoT devices has transformed the way we interact with physical systems, enabling efficient remote automation and monitoring. A key advancement in this space is Edge AI, which brings AI computation closer to the data source—typically at the network edge—to enhance  responsiveness and reduce latency \cite{SINGH202371}. However, Edge AI solutions are inherently resource-constrained, driving research efforts toward optimizing hardware efficiency. While hardware acceleration has received significant attention, the methodology behind designing ISA extensions remains an equally important yet often underexplored aspect. Many existing works introduce ISA extensions based on assumed computational hotspots in DNNs such as matrix multiplications or rely on heuristic, domain-specific motivations without empirical profiling. For example, Xvpfloat~\cite{10488759} introduces a RISC-V ISA extension designed for dynamically variable and extended precision floating-point computations. The extension supports significant sizes up to 512 bits and includes features like specialized indexed loads/stores and hardware-assisted prefetching. However, this design is driven by the need for efficient hardware support in scientific computing, not by application profiling. Similarly, Convex~\cite{10.1145/3675018.3675029} proposes a RISC-V ISA extension aimed at enhancing the performance of convolution operations on microcontrollers. The extension includes bit concatenation, weight storage optimization, and activation value reuse, yet these are guided by high-level architectural goals rather than profiling data. Authors in \cite{10.1145/3587135.3592168} describe Confidential Virtual Machine Extension (CoVE) for RISC-V, aiming to provide hardware support for confidential computing. The proposed ISA and non-ISA extensions focus on security features necessary for trusted execution environments. Here too, the design decisions are driven by security requirements rather than application profiling.

Another often overlooked consideration is the ease of programming and software integration, which plays a crucial role in the practical adoption of these extensions. Despite advances in AI hardware, the lack of automated and user-friendly software solutions remains a significant barrier to widespread adoption. Existing solutions either focus on translating high-level domain-specific languages (DSLs) like PyTorch and TensorFlow into low-level code, or offloading the programming responsibility entirely to the user while focusing solely on hardware efficiency. However, very few solutions provide a complete, end-to-end framework that seamlessly integrates AI models with specialized hardware. For instance, BARVINN \cite{10.1145/3566097.3567872}, a RISC-V-based DNN accelerator, processes ONNX inputs and generates executable command streams for a RISC-V controller. However, its software framework lacks flexibility, as it does not support residual connections, limiting compatibility with models like ResNet. RISC-VTF \cite{9658643} extends the RISC-V instruction set to improve transformer model execution through custom matrix operations and activation functions, yet the absence of compiler support forces users to write low-level RISC-V assembly. Similarly, Eyeriss \cite{eyeris}, designed as a co-processor to accelerate CNN workloads,
requires a robust software stack to efficiently offload compute-intensive tasks. Other solutions, such as the LSTM accelerator by Kadetotad et al. \cite{tops}, optimize speech recognition workloads but require substantial modifications in quantization, compression, and memory structure, making them unsuitable for off-the-shelf pretrained models. Another RISC-V-based processor \cite{7864441} for IoT endpoints  introduces DSP extensions for energy-efficient filtering, convolutions, and simple ML operations. While the compiler toolchain supports vectorization and hardware loops, it lacks an automated model conversion pipeline from high-level AI frameworks to embedded C, requiring manual coding efforts.

\par The challenge of limited software accessibility is also evident in model-specific instruction-based accelerators. XpulpNN \cite{xpulp} enhances a RISC-V core with SIMD operations for quantized neural networks (QNN), achieving significant speedups. However, details on its programming interface remain sparse. ZeroRiscy \cite{cnnrisc} accelerates CNN workloads but requires developers to write RISC-V assembly code to leverage custom ISA extensions. In another example, Aness et al. \cite{10737828} optimize transformer-based models for constrained devices by porting ARM Keyword Spotting Transformer \cite{Berg_2021}, reducing model size by 369$\times$ with only a 10\% accuracy drop, yet this relies on manual C-code modifications without an automated pipeline.
\par A broader challenge arises when designing general-purpose instruction-based accelerators, where software-hardware co-design is crucial for usability. FlexACC \cite{9516466} addresses model-
accelerator incompatibilities by introducing ISA extensions for MLP, CNN, LSTM, GCN, and transformer models, achieving a 216$\times$ CNN speedup. However, its programming model remains low-level and requires C-like coding. Similarly, \cite{10.1145/3665283.3665342} explores an ISA extension that optimizes memory access patterns in AI applications, reducing power consumption and code size without significant area overhead. However, compiler updates are necessary for full utilization, and support for non-consecutive memory accesses remains limited. Efforts have also been made to distribute computation between IoT devices and edge servers to reduce latency and enhance privacy. For example, \cite{10909993} presents a framework that integrates pruning and early exit strategies with FPGA-based offloading, achieving a 1.6$\times$ latency reduction and 3.9$\times$ power efficiency improvement. However, its complexity makes adaptation to different AI models challenging. Compute-in-memory architectures \cite{10463613} introduce dedicated RISC-V ISA extensions for switching operational modes but lack an end-to-end flow that supports high-level AI model inputs as it expects the software program to be loaded to bootRAM from external flash with weight parameters already stored into multi-bank eMRAM. AI-RISC \cite{verma2022ai} represents a more integrated approach, co-designing hardware, ISA, and software to extend RISC-V for Edge AI acceleration. By embedding AI functional units into the RISC-V pipeline, it enables seamless execution of both AI and non-AI tasks. However, AI-RISC requires modifications to the TVM compiler to recognize its intrinsics, making adaptation to different models cumbersome. In contrast, our work eliminates the need for such modifications, providing a streamlined, adaptable, and automated solution for bridging high-level AI frameworks with edge hardware. By addressing the gap in software accessibility, we introduce a framework that seamlessly integrates AI models with resource-constrained accelerators, facilitating efficient and user-friendly deployment of Edge AI solutions.
\par This paper presents an automated workflow that converts any Python-based CNN model into a bare metal implementation for a RISC-V CPU enhanced with the proposed ISA extensions, optimizing both speed and energy efficiency.
For evaluation, the design was implemented on a Xilinx ZCU104 FPGA and the performance is benchmarked on C implementations of pretrained DNNs.  The latency required to perform model inference is compared between the baseline RISC-V core and the same core with added custom ISA extensions.
The main contributions of this paper (and its comparison with some related works in Table \ref{related_works_table}) can be summarized as follows:
\begin{itemize}
\item An end-to-end flow showcasing how Python-based DNN models from high-level frameworks like PyTorch/TensorFlow are translated to C and profiled to develop a model class-specific custom RISC-V which enables bare metal programming. 
\item A RISC-V core with a custom ISA generated with the proposed MARVEL flow for lightweight CNNs which achieves an inference acceleration of up to \textbf{2$\times$}.
\item A tool flow based on ASIP Designer \cite{asip_designer}, to facilitate an accelerated modeling and development of the identified ISA extensions.  The flow described in this paper tracks the design of these extensions from Python models to final hardware implementation.
\item Benchmarking of DNN models to quantify the performance of the custom RISC-V core against the baseline. FPGA demonstration of the identified extensions with a reduction in energy per inference by up to 2$\times$ for models such as LeNet-5*\footnote{Hand-coded C implementation of LeNet-5-like model}, MobileNetV1, ResNet50, VGG16, MobileNetV2 and DenseNet121.
\end{itemize}

\section{Methodology}

\begin{figure*}
    \centering
    \includegraphics[width=1\linewidth]{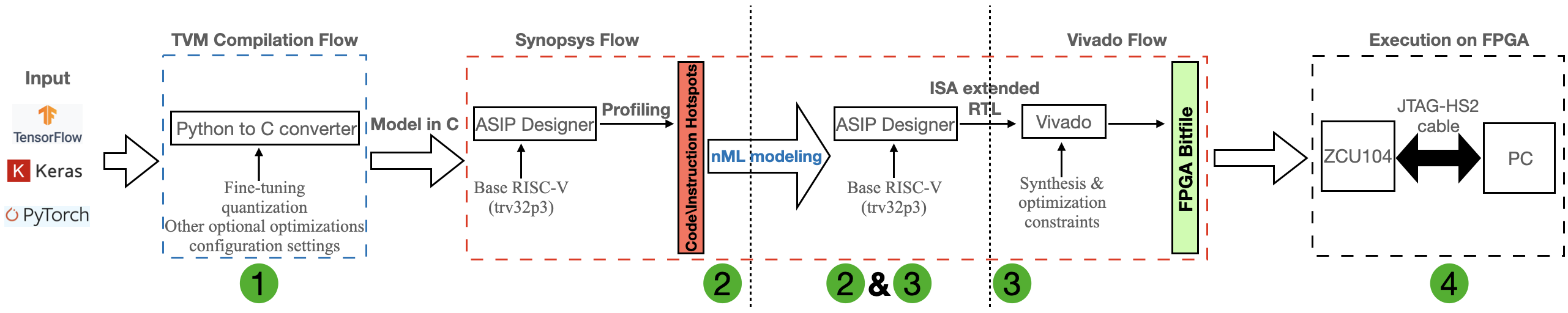}
    \caption{MARVEL: An end-to-end framework to generate model-class specific custom RISC-V for DNN acceleration}
    \label{top_level_flow_block_diagram}
\end{figure*}

\begin{figure*}
    \centering
    \includegraphics[width=1\linewidth]{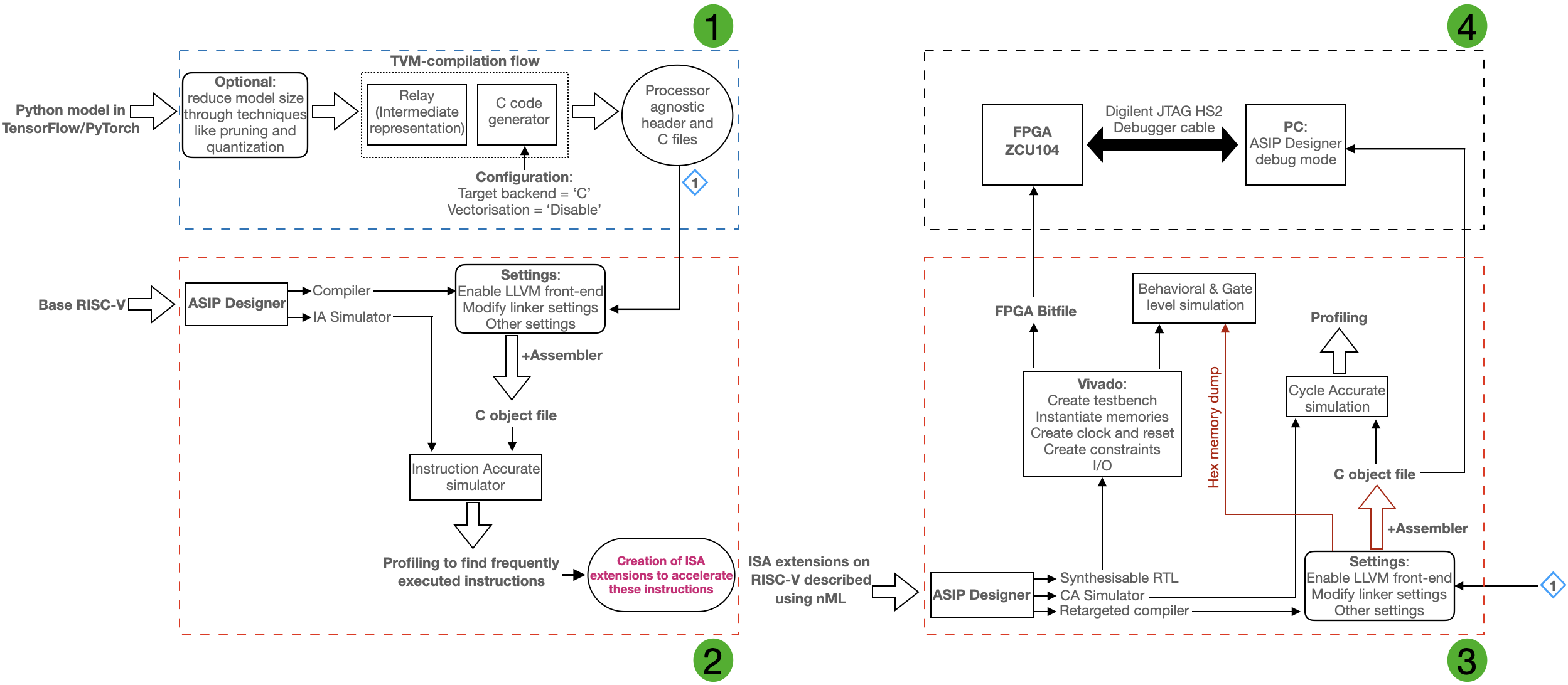}
    \caption{MARVEL: Detailed flow diagram}
    \label{top_level_flow}
\end{figure*}

The trv32p3 processor core~\cite{trv32p3} from Synopsys is used as a starting point for the development of an Application Specific Instruction set Processor (ASIP). trv32p3 has a 32-bit wide data path and a three-stage pipeline. This core implements the RV32IM ISA that includes integer instructions and hardware support for integer multiplication, division and remainder. Additionally,  on-chip debugging (OCD) for JTAG access and memory was integrated before compiling with ASIP designer to generate HDL files for FPGA implementation.
The design was synthesized targeting an AMD ZCU104 development platform.
The extended core was tested with quantised standard DNN models' C codes obtained through the proposed workflow as described further. The complete process, from converting a Python-based model to executing it on the extended RISC-V core implemented on an FPGA, is illustrated in Fig~\ref{top_level_flow_block_diagram} and~\ref{top_level_flow}

\subsection{Workflow to generate C code}
\label{proposed_flow}

This section outlines the methodology for converting an AI model, implemented in Python using high-level deep learning frameworks such as TensorFlow, into C code suitable for compilation with Synopsys' ASIP Designer tools. The target architecture in this work is the RISC-V-based trv32p3 processor and its ISA extended variants. To enable this conversion, we leverage the TVM machine learning compiler\cite{10.5555/3291168.3291211}, an open source framework designed to optimize and deploy deep learning models on a variety of hardware backends. Notably, there are other commercially available frameworks that also enable this translation \cite{edgecortix},\cite{iree}. The translation process begins with the conversion of the high-level Python-based model into Relay, an intermediate representation (IR) that is hardware-agnostic and optimized for computational graphs. Relay serves as an abstraction layer, enabling transformations and optimizations independent of the underlying hardware target. Once the model is represented in Relay IR, TVM applies a series of optimization passes to enhance computational efficiency before generating source code tailored for the specified backend. In this work, we configure TVM to target generic C, producing C-based computational kernels that can be compiled and executed on any processor that supports C compilation. This approach facilitates portability and enables seamless integration of AI workloads on RISC-V based embedded systems. The following subsections detail the step-by-step procedure for generating C code using TVM, from model parsing to final code generation. Additionally, an open-source repository has been made available on GitHub\footnote{\url{https://github.com/ajaymahendersingh/MARVEL}}, containing example scripts and reference implementations for converting DNN models into C code suitable for RISC-V deployment.
\vspace{-6mm}
\subsubsection{Environment Setup}
To begin, all required dependencies and software packages must be installed. The models used in this work are pre-trained keras deep learning architectures, which are subsequently fine-tuned for the target classification task. Five deep learning models—MobileNetV1, ResNet50, VGG16, MobileNetV2, and DenseNet121—are fine-tuned using the StanfordCars\cite{6755945} and COCO\cite{cocodataset} datasets. The input images are standardized to dimensions of 64×64×3 (height, width, channels), and the classification task is simplified to a binary problem: distinguishing between "Car" and "Not Car." In addition, a modified LeNet-5 model is employed to classify 28×28 grayscale images of handwritten digits into one of the ten-digit classes. 
\vspace{-6mm}
\subsubsection{Model Training Using Transfer Learning}
This step is optional and not part of the main workflow, though it is recommended. The proposed approach is designed for IoT-embedded systems with limited resources. For these systems, minimizing network size is crucial for compatibility. Therefore, in this step, we fine-tune the standard models to focus on two target classes. To adapt the pre-trained models for the target application, transfer learning is employed. The models are re-trained for a limited number of epochs until the validation accuracy reaches an acceptable level. The dataset is split into an 80:20 ratio for training and validation. To accommodate the binary classification task, the final fully connected layer of the original network is removed and replaced with a dense layer containing two neurons, corresponding to the two output classes. %

\subsubsection{Model Size Optimization via Quantization}
This step is optional but included in the proposed flow to further reduce the network size. Once the model is trained, it exists as a TensorFlow-based DNN tailored to the car recognition task. However, its memory footprint remains prohibitively large for deployment on embedded platforms with constrained memory resources. For instance, MobileNetV1, one of the models used in this work, contains 216,000 parameters, each represented as a 32-bit floating-point number in the native model. This translates to very high storage and computational requirements, making them non-viable for deployment on edge platforms. To mitigate this, post-training quantization is applied to reduce model size and improve inference efficiency. The TensorFlow Lite framework~\cite{tflite} is leveraged to convert the model’s floating-point parameters into 8-bit integers, significantly reducing both memory consumption and computational overhead without substantial accuracy degradation\cite{LIANG2021370}.
\vspace{-6mm}
\subsubsection{Conversion of Python Model to Bare-Metal C}
Up to this stage, TensorFlow has been used to construct and train the DNN model. For deployment on the trv32p3 RISC-V processor, the model is first converted into C code \footnote{This does not include the hand-coded LeNet-5* implementation, as it is already written in C.} for compilation with ASIP Designer. This conversion is performed using TVM compiler as shown in Fig.~\ref{top_level_flow}. TVM translates high-level Python-based AI models into optimized C code, tailored for the specified target hardware. Since the RISC-V processor used in this work operates without an operating system (bare metal environment), we configured TVM to generate generic C code.
\vspace{-6mm}
\subsubsection{TVM-generated C Application}
Once the model is successfully compiled with TVM, the generated C source files are stored in a structured directory. The key output files are:

\begin{itemize}
    \item Header file defining structs for inputs and outputs → ./codegen/host/include/tvmgen\_default.h
    \item C source files containing the compiled neural network weights and implementation → ./codegen/host/src/default\_lib0.c , ./codegen/host/src/default\_lib1.c
\end{itemize}
In addition to the compiled source files, sample input images are provided for validation purposes.
No modifications are required for the generated C source files. To integrate the model into an application, a custom main function must be implemented to invoke the necessary TVM-generated functions for inference.

\subsection{ASIP Designer}
The trv32p3 ISA is extended using ASIP Designer (version W-2024.12). ASIP Designer \cite{asip_designer} is a retargetable compiler, simulation, and hardware generation environment which uses the special-purpose modeling languages nML and PDG (Primitives Definition and Generation).  nML is a high-level language for modeling processor architectures. An nML description of a target processor can be converted into a hardware description with the use of the Synopsys Go\textsuperscript{\textregistered} nml-to-hdl compiler~\cite{nml-to-hdl_compiler}. The primitives used by an nML processor description are described in the separate PDG language. It is with this language that the specific processor functionality, I/O behaviour and control behaviour are described. The custom instructions described in the next section are added to the existing nML model for the trv32p3 processor, with the specific functionality defined in a separate PDG description.

\subsection{ISA extensions}
\label{ISA_extensions}
\begin{figure*}[ht!]
  \centering
  \begin{minipage}{0.33\textwidth}
    \centering
    \includegraphics[width=\linewidth]{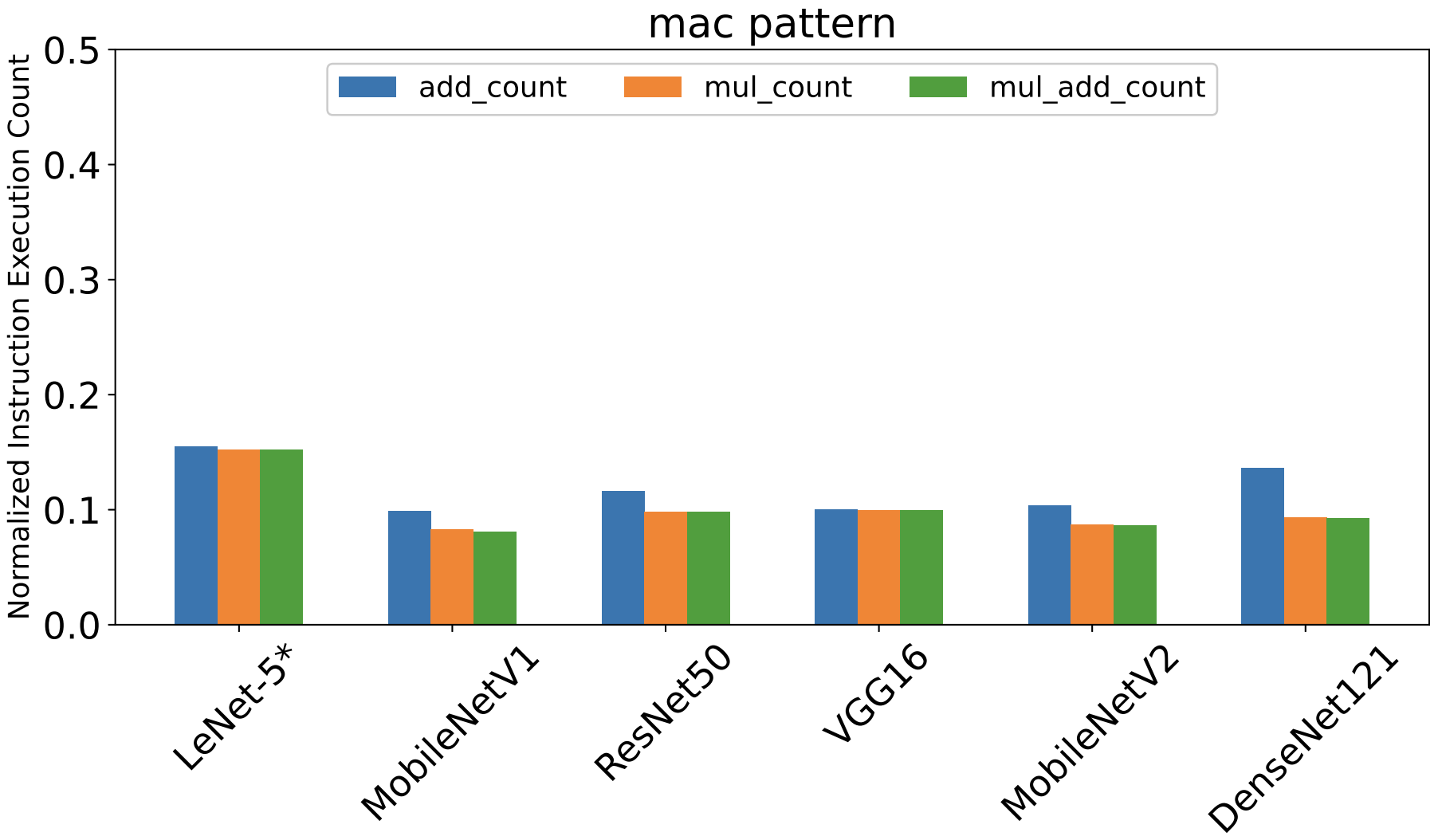}
  \end{minipage}%
  \begin{minipage}{0.33\textwidth}
    \centering
    \includegraphics[width=\linewidth]{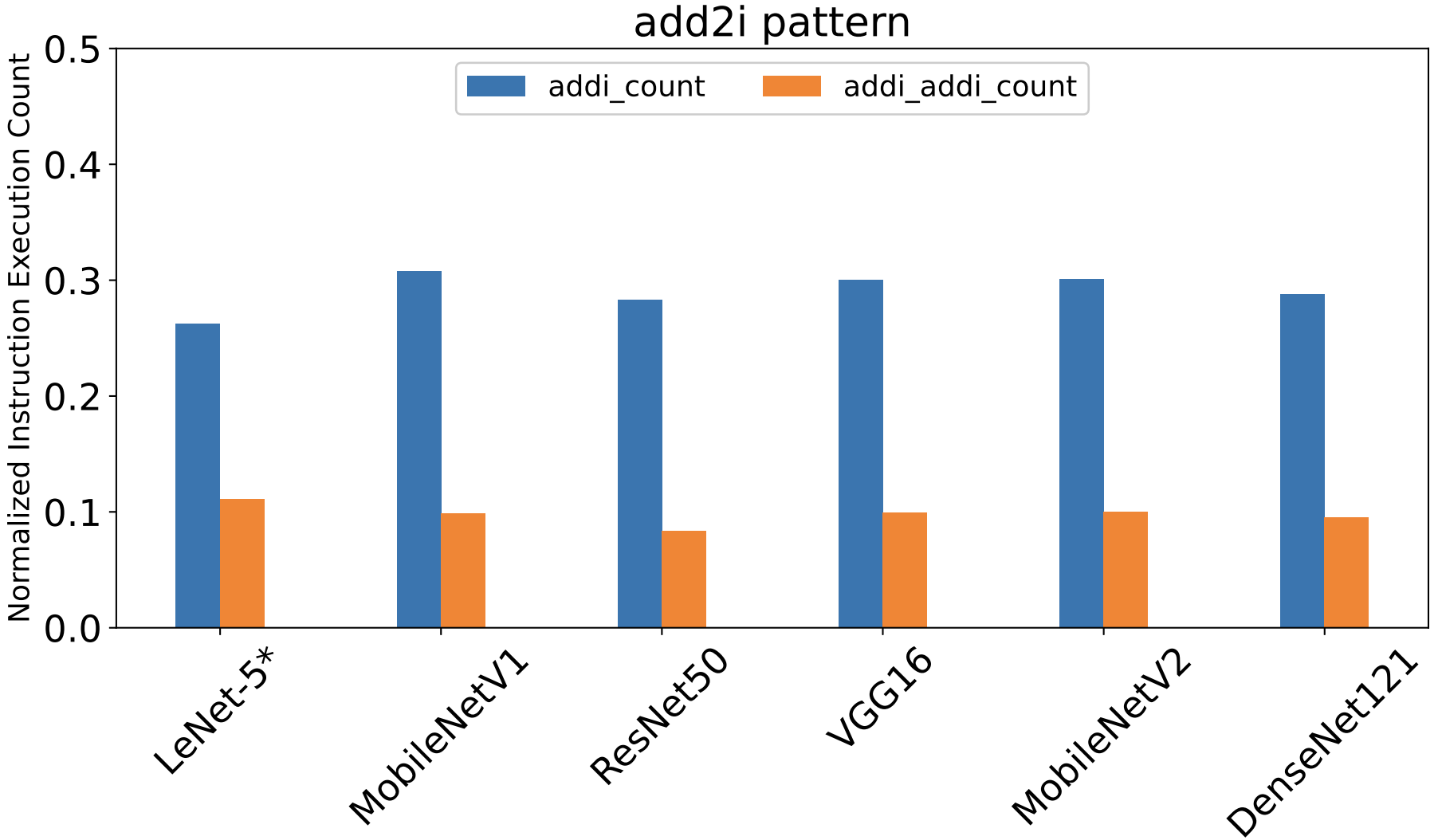}
  \end{minipage}
  \begin{minipage}{0.33\textwidth}
    \centering
    \includegraphics[width=\linewidth]{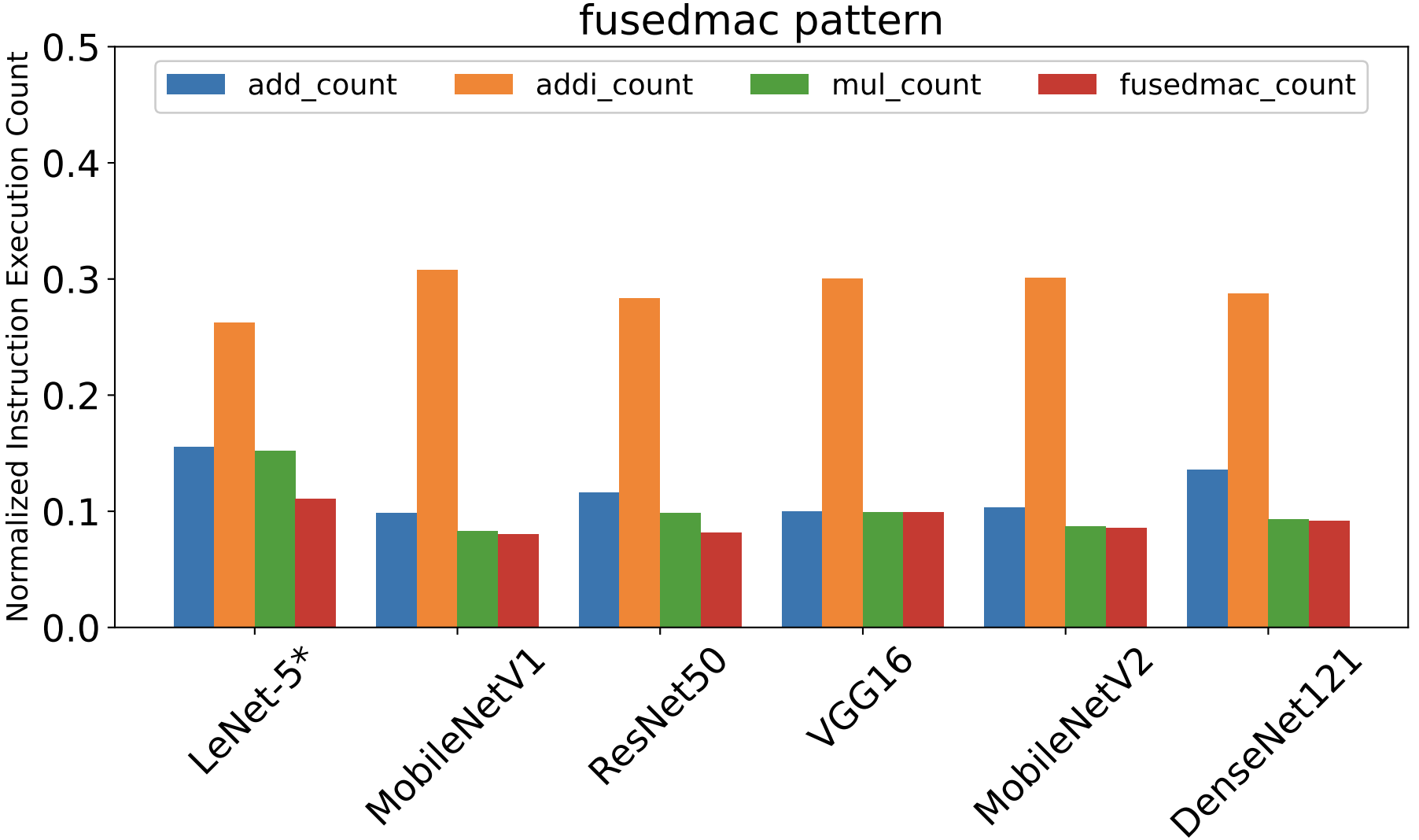}
  \end{minipage}%
  \caption{Frequently executed patterns identified through profiling of generated C code on baseline RISC-V}
  \label{pattern_count_plots}
\end{figure*}

Traditionally, ISA extensions are introduced to address known computational bottlenecks, often without a comprehensive analytical foundation. For example, LiteAIR5 \cite{10257058} proposes ISA extensions targeting matrix-matrix multiplication, a well-known bottleneck in AI workloads. In this study, we took a more data-driven approach by profiling (step 2 in Fig.~\ref{top_level_flow_block_diagram}) C code generated through the proposed flow for MobileNetV1 on a baseline RISC-V architecture to systematically identify performance bottlenecks. Using Synopsys tools, we analyze execution patterns to pinpoint compute-intensive operations that could benefit from ISA extensions. Specifically, we enable instruction profiling in a instruction-accurate simulator, capturing execution counts for each instruction. We then sort and analyze the most cycle-intensive instructions, identifying key compute bottlenecks that can be addressed through ISA extensions. Additionally, it would be valuable to conduct a similar analysis using RISC-V’s open-source toolchain and Spike simulator, as demonstrated in \cite{10752449}, to obtain detailed instruction coverage metrics for specific RISC-V extensions. Our profiling revealed a set of frequently executed instruction patterns that, when optimized, could significantly enhance performance. To validate the applicability of our proposed extensions, we extended our profiling to other models within the same class. The results confirmed that the identified patterns were not model-specific but rather class-specific, making them broadly applicable. Fig~\ref{pattern_count_plots} shows the normalised count of the number of times these patterns are executed in the corresponding model, while Table \ref{pattern_count_plots_desc} provides definitions for the legends used in this figure. These patterns were then optimized through the proposed ISA extensions utilizing the custom opcodes available in RISC-V as outlined in Table \ref{inst_encode}. The following subsections provide a detailed discussion of these extensions, with Table \ref{processor_versions} defining the terminology used to describe different processor versions as they evolve with custom instructions.
\vspace{-3mm}
\begin{table}[H]
    \centering
    \scriptsize
    \caption{Terminology for different RISC-V variants}
    \label{processor_versions}
    \begin{tabular}{cl}
        \toprule
        \textbf{Processor Version} & \textbf{Description} \\ 
        \midrule
        v0 & Baseline RISC-V processor (trv32p3) \\  
        v1 & \verb|mac| extension enabled on v0 \\  
        v2 & \verb|add2i| extension enabled on v1 \\  
        v3 & \verb|fusedmac| extension enabled on v2 \\  
        v4 & Zero-overhead hardware loops (\verb|zol|) extension enabled on v3 \\  
        \bottomrule
    \end{tabular}
\end{table}
    
\begin{table}[h]
  \centering
  \scriptsize
  \caption{Description of instruction patterns in Fig. \ref{pattern_count_plots}}
  \label{pattern_count_plots_desc}
  \begin{tabular}{llp{4.8cm}}
    \toprule
    Pattern & Metric & Description \\ 
    \midrule
    \multirow{3}{*}{\texttt{mac}} 
        & add\_count & Number of times the \verb|add| instruction occurs \\  
        & mul\_count & Number of times the \verb|mul| instruction occurs \\  
        & mul\_add\_count & Number of times the \verb|mul| and \verb|add| instructions occur consecutively \\  
    \midrule
    \multirow{2}{*}{\texttt{add2i}}  
        & addi\_count & Number of times the \verb|addi| instruction occurs \\  
        & addi\_addi\_count & Number of times two \verb|addi| instructions occur consecutively \\  
    \midrule
    \multirow{4}{*}{\texttt{fusedmac}}  
        & add\_count & Number of times the \verb|add| instruction occurs \\  
        & addi\_count & Number of times the \verb|addi| instruction occurs \\  
        & mul\_count & Number of times the \verb|mul| instruction occurs \\  
        & fusedmac\_count & Number of times the \verb|mul|, \verb|add|, and two \verb|addi| instructions occur consecutively \\  
    \bottomrule
  \end{tabular}
\end{table}

\begin{table}[ht!]
    \centering
    \caption{Instruction opcode encoding}
    \includegraphics[width=\linewidth]{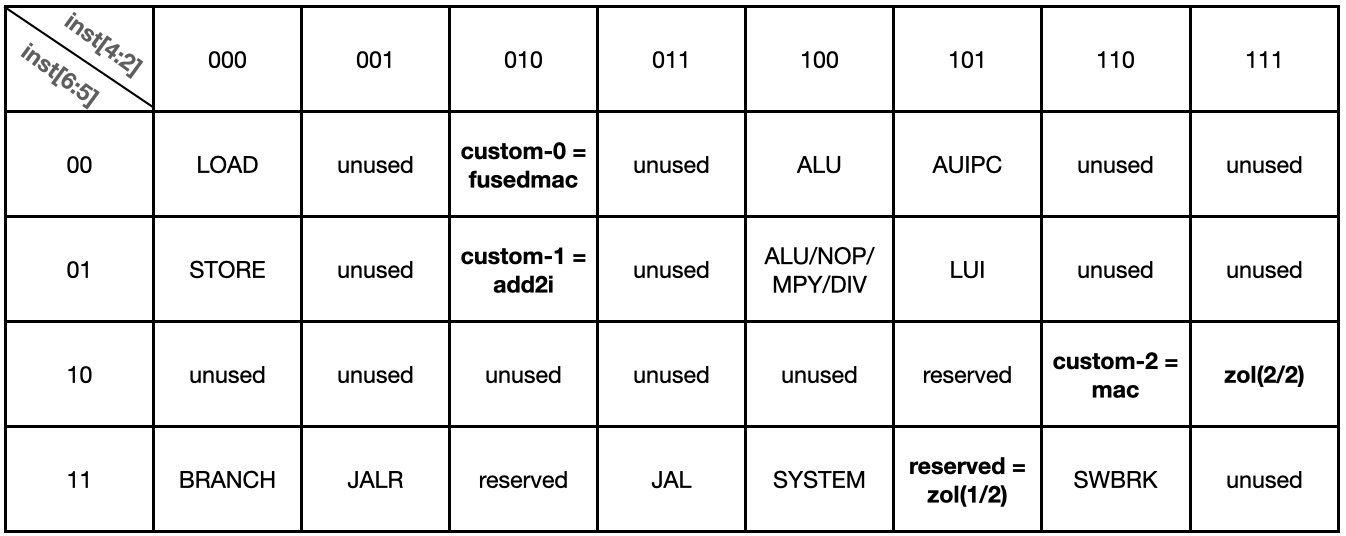}
    \label{inst_encode}
\end{table}
\vspace{-10mm}
\subsubsection{mac}
 The instruction ``\verb|mul rd, rs1, rs2|'' performs a 32-bit multiplication, followed by the instruction ``\verb|add rd, rd, rs1|'' which accumulates the result through addition. This pair of instructions can be replaced by a single custom ``\verb|mac rd, rs1, rs2|''. This operation performs the same operation in half the number of clock cycles. This is the most common operation in all AI/ML workloads\cite{VERMA2022100742}. In this work, we fix the registers (``\verb|rd = x20, rs1 = x21, rs2 = x22|''), which enables us to combine mac with another custom instruction in subsequent steps. Hardcoding the source/sink registers also reduces the hardware area overhead incurred by the processor version, where both these custom instructions co-exist. The trade-off is reduced flexibility in register allocation, though this had minimal impact in practice due to ample general-purpose registers.
 This extension uses the CUSTOM 2 opcode. Listing~\ref{mac_listing} shows the assembly syntax for this instruction, and Table~\ref{mac_inst_decoding} shows its decoding scheme.

\lstset{
    language={[x86masm]Assembler},
    basicstyle={\ttfamily\small},
    keywordstyle=\bfseries,
    morekeywords={mac, add2i, fusedmac, zol},
    commentstyle=\itshape,
    numbers=left,
    numberstyle=\tiny,
    stepnumber=1,
    numbersep=5pt,
    tabsize=4,
    frame=single,
    columns=fullflexible,
    showstringspaces=false
}
\begin{figure}[h]
\centering
\begin{minipage}{0.96\linewidth}
\begin{lstlisting}[caption={Assembly format of the custom mac instruction}, label={mac_listing}, basicstyle=\small]
    ;rd = rd + rs1*rs2
    mac   
    ;rd = x20, rs1 = x21, rs2 = x22
\end{lstlisting}
\end{minipage}
\end{figure}

\begin{table}[h]
    \centering
    \caption{Instruction decoding: mac}
    \includegraphics[width=\linewidth]{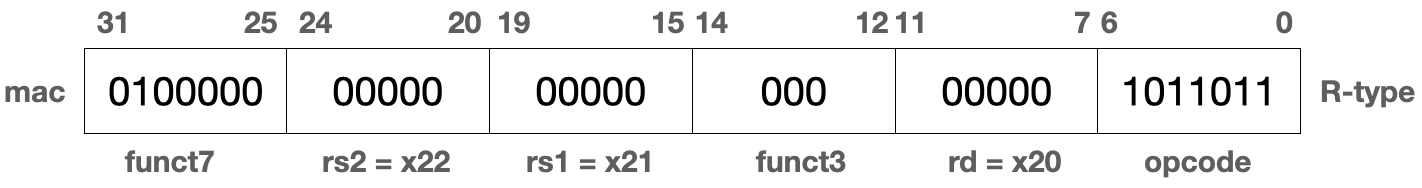}
    \label{mac_inst_decoding}
\end{table}
\subsubsection{add2i}
\begin{figure*}[h]
  \centering
  \begin{minipage}{0.163\linewidth}
    \centering
    \includegraphics[width=\linewidth, height=3cm]{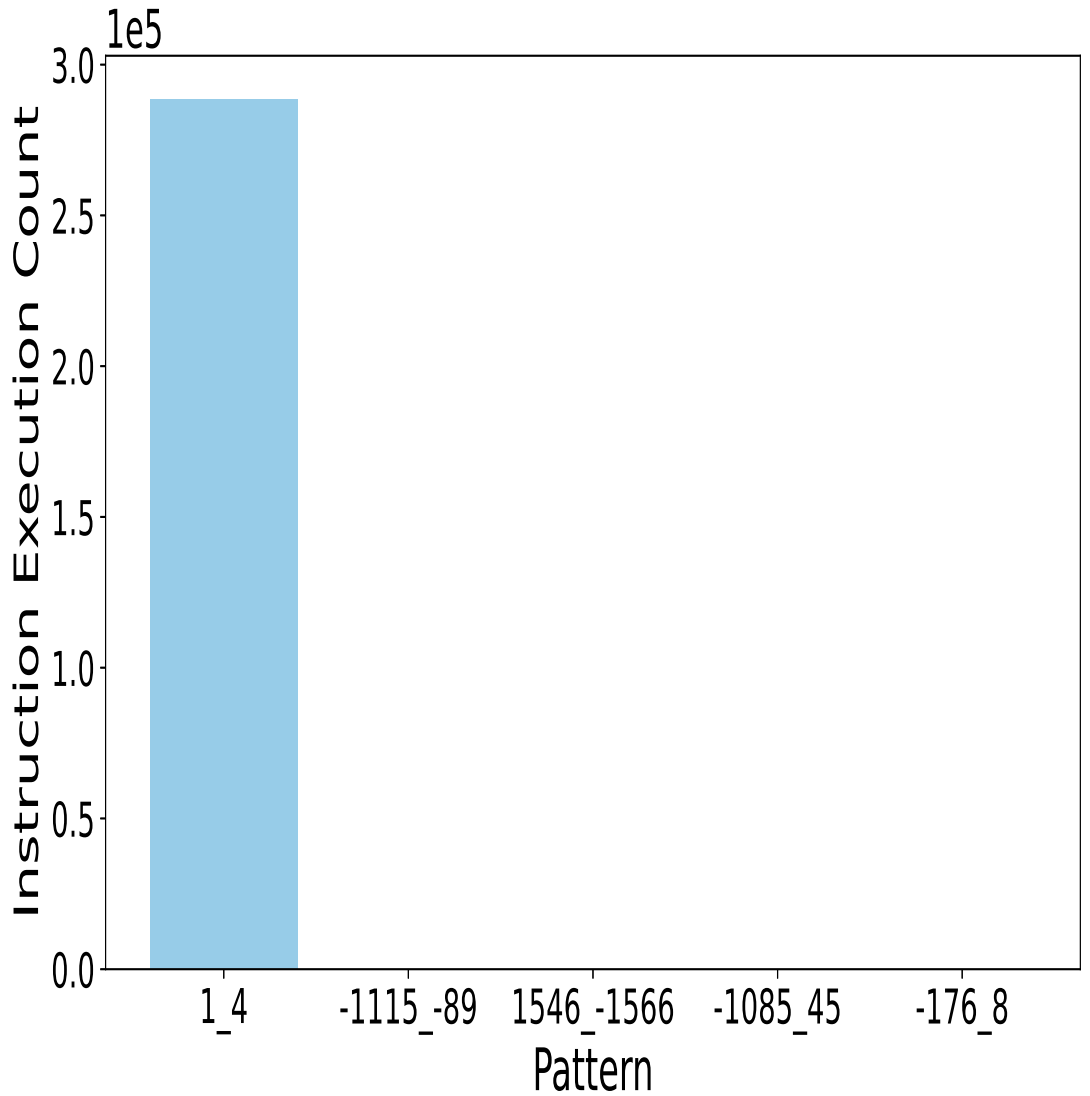}
    \caption*{\footnotesize{LeNet-5*}}
  \end{minipage}%
  \begin{minipage}{0.163\linewidth}
    \centering
    \includegraphics[width=\linewidth, height=3cm]{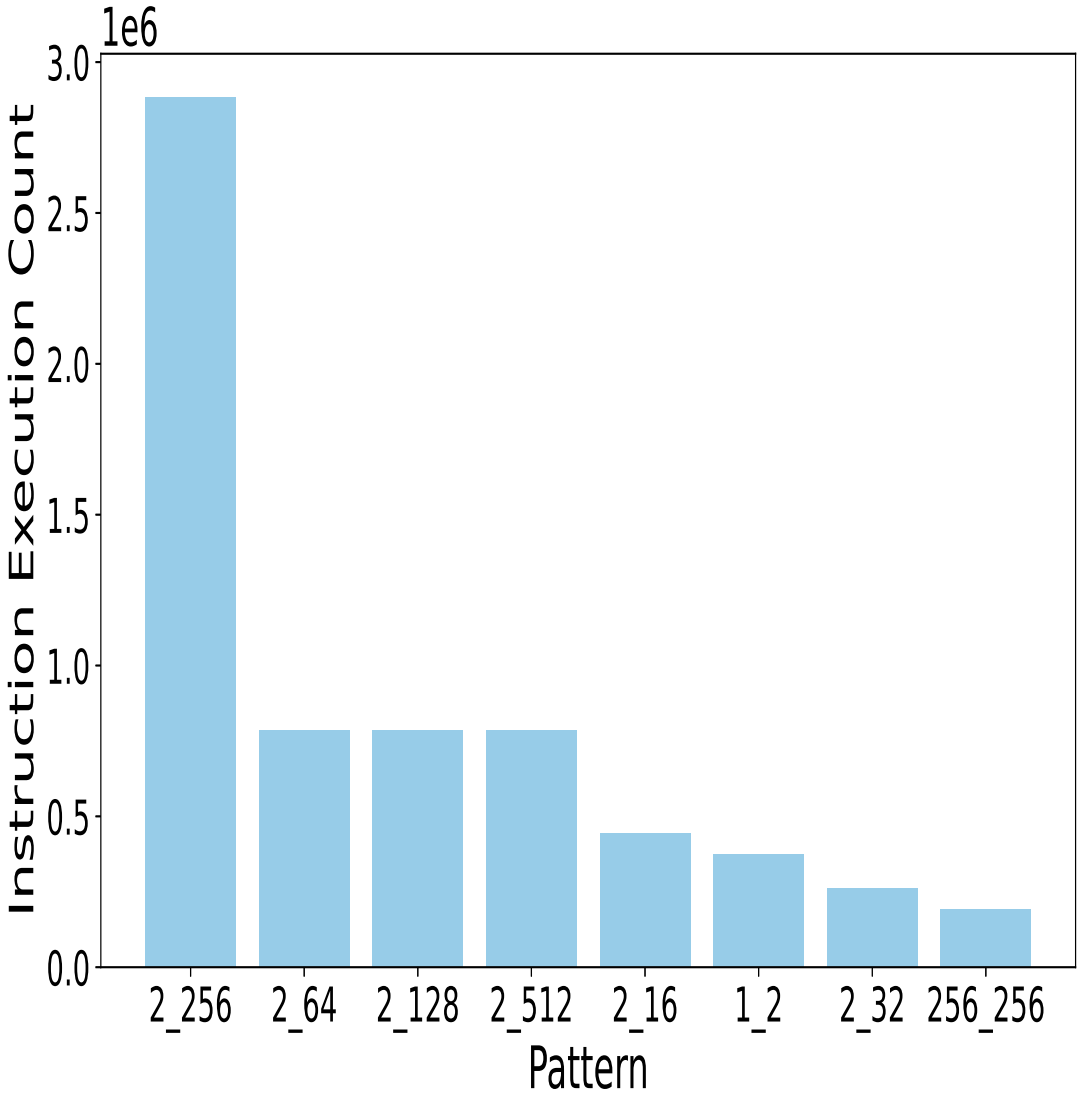}
    \caption*{\footnotesize{MobileNetV1}}
  \end{minipage}
  \begin{minipage}{0.163\linewidth}
    \centering
    \includegraphics[width=\linewidth, height=3cm]{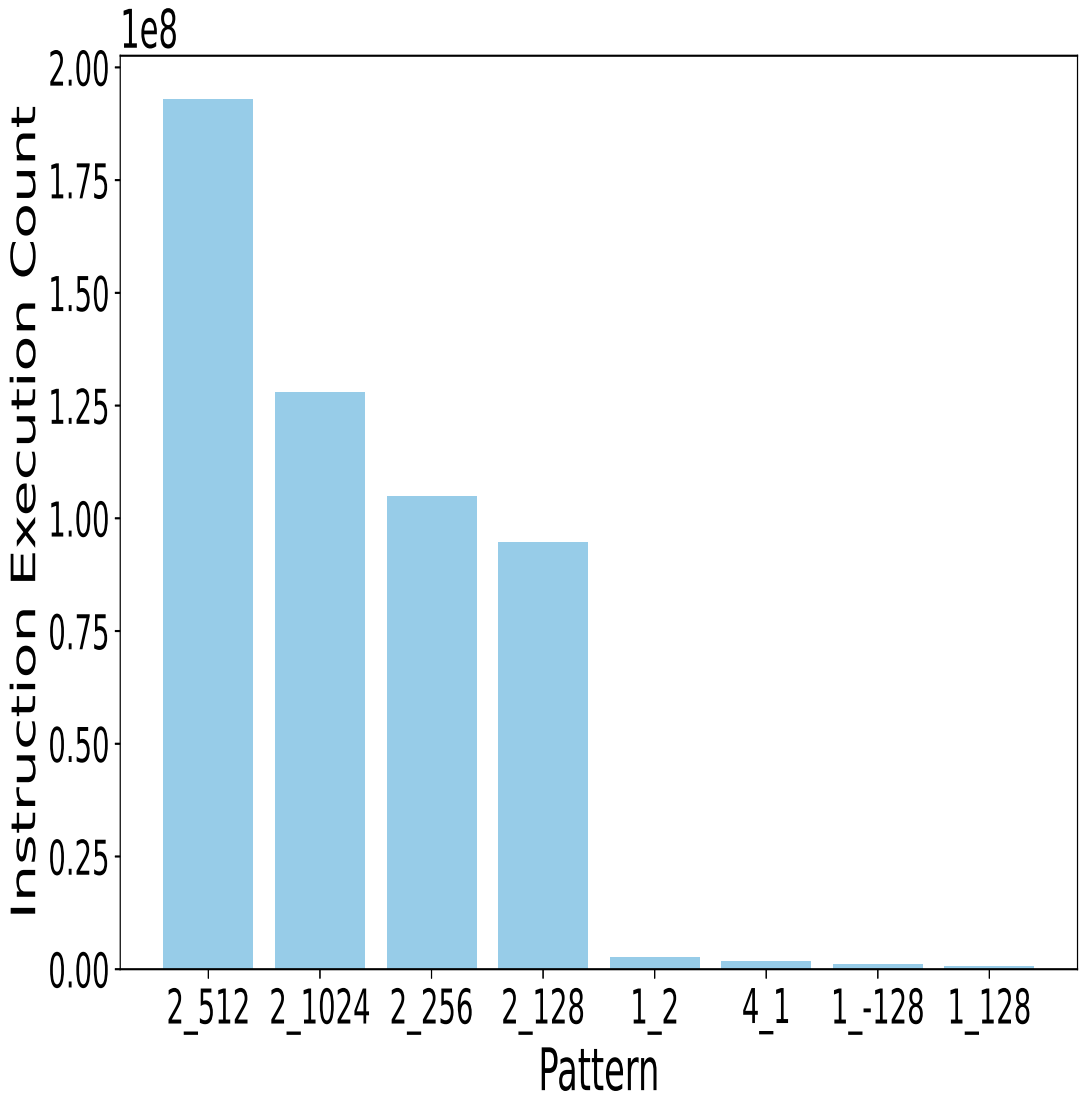}
    \caption*{\footnotesize{ResNet50}}
  \end{minipage}%
  \begin{minipage}{0.163\linewidth}
    \centering
    \includegraphics[width=\linewidth, height=3cm]{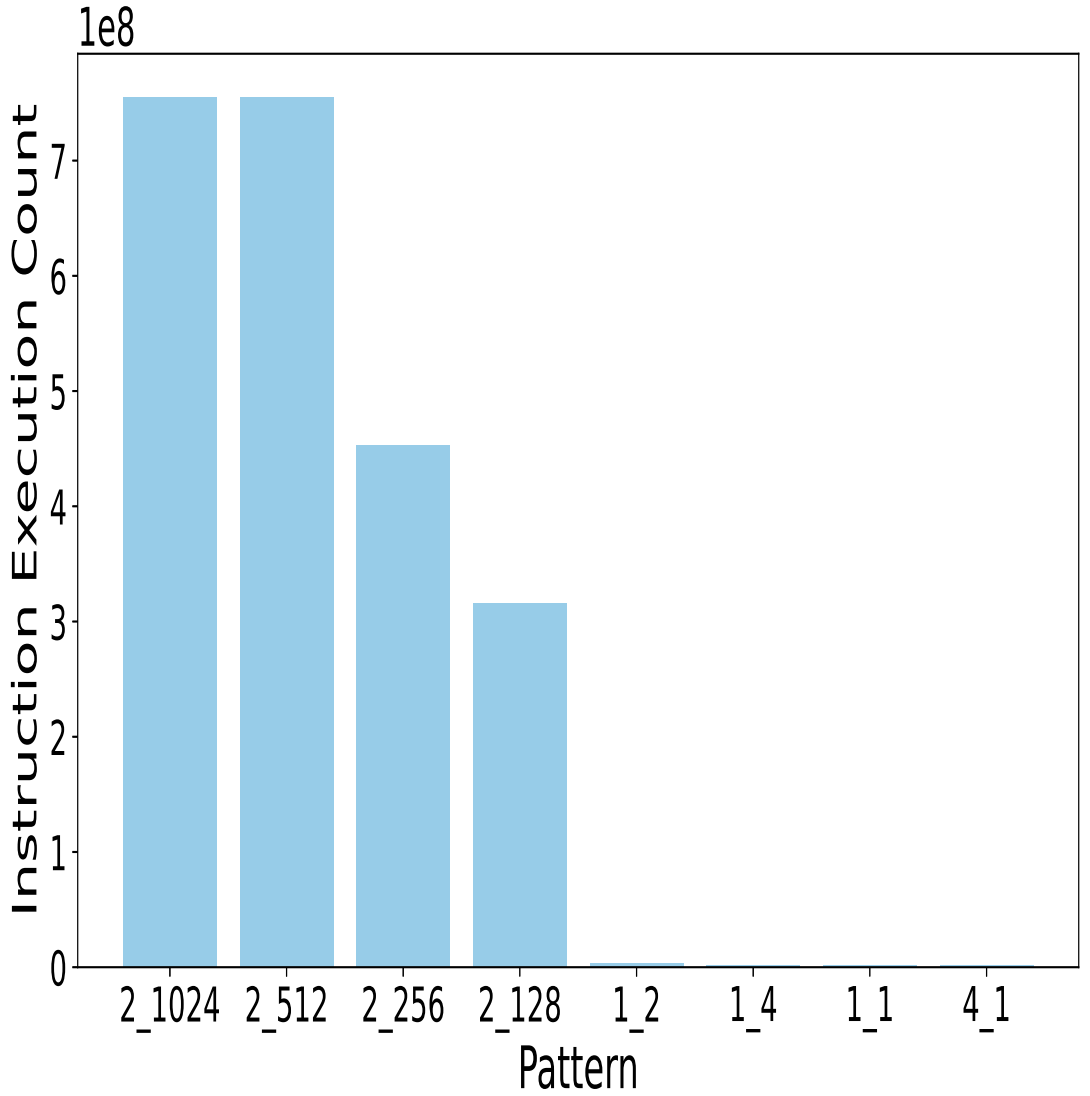}
    \caption*{\footnotesize{VGG16}}
  \end{minipage}
  \begin{minipage}{0.163\linewidth}
    \centering
    \includegraphics[width=\linewidth, height=3cm]{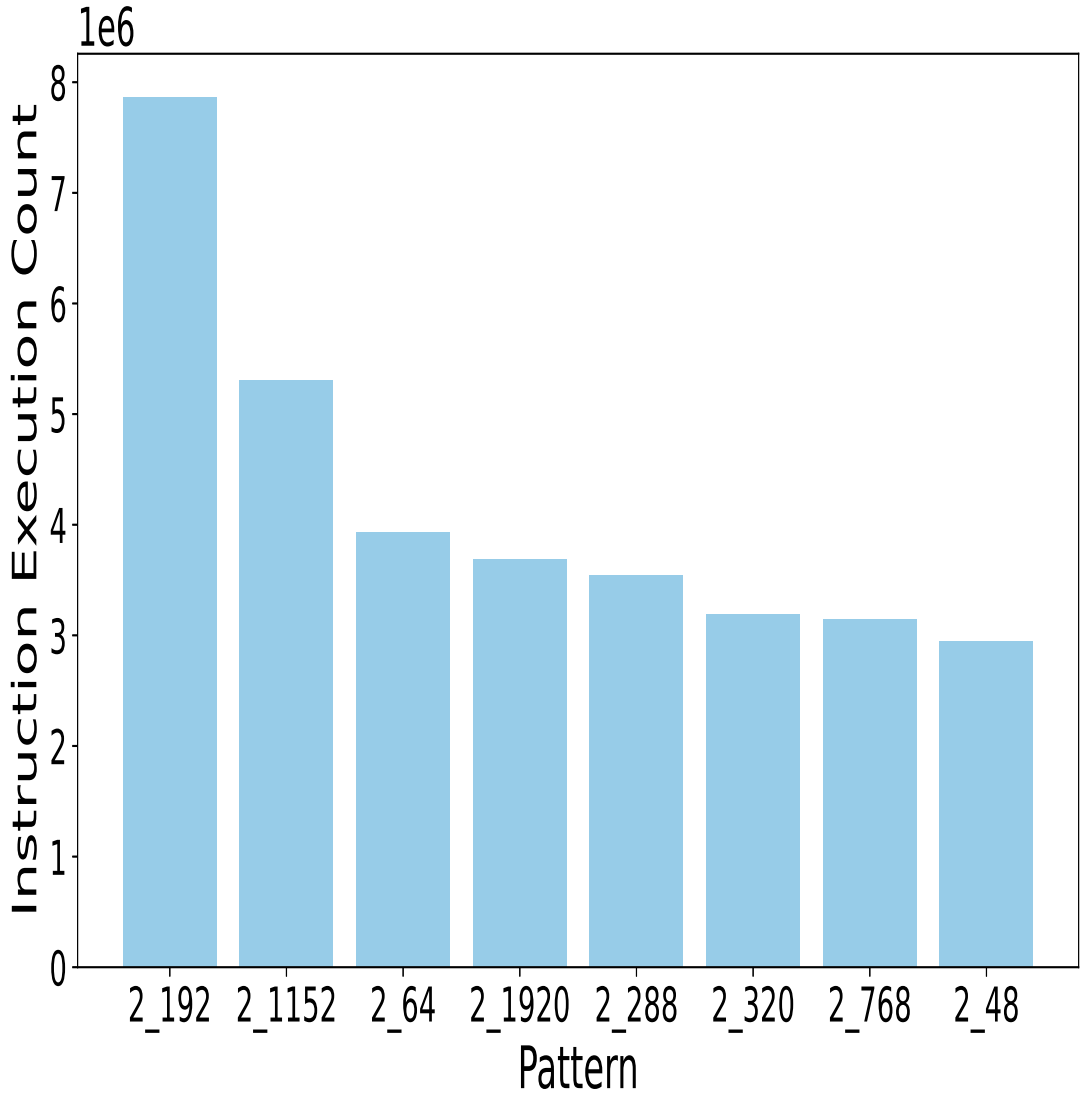}
    \caption*{\footnotesize{MobileNetV2}}
  \end{minipage}%
  \begin{minipage}{0.163\linewidth}
    \centering
    \includegraphics[width=\linewidth, height=3cm]{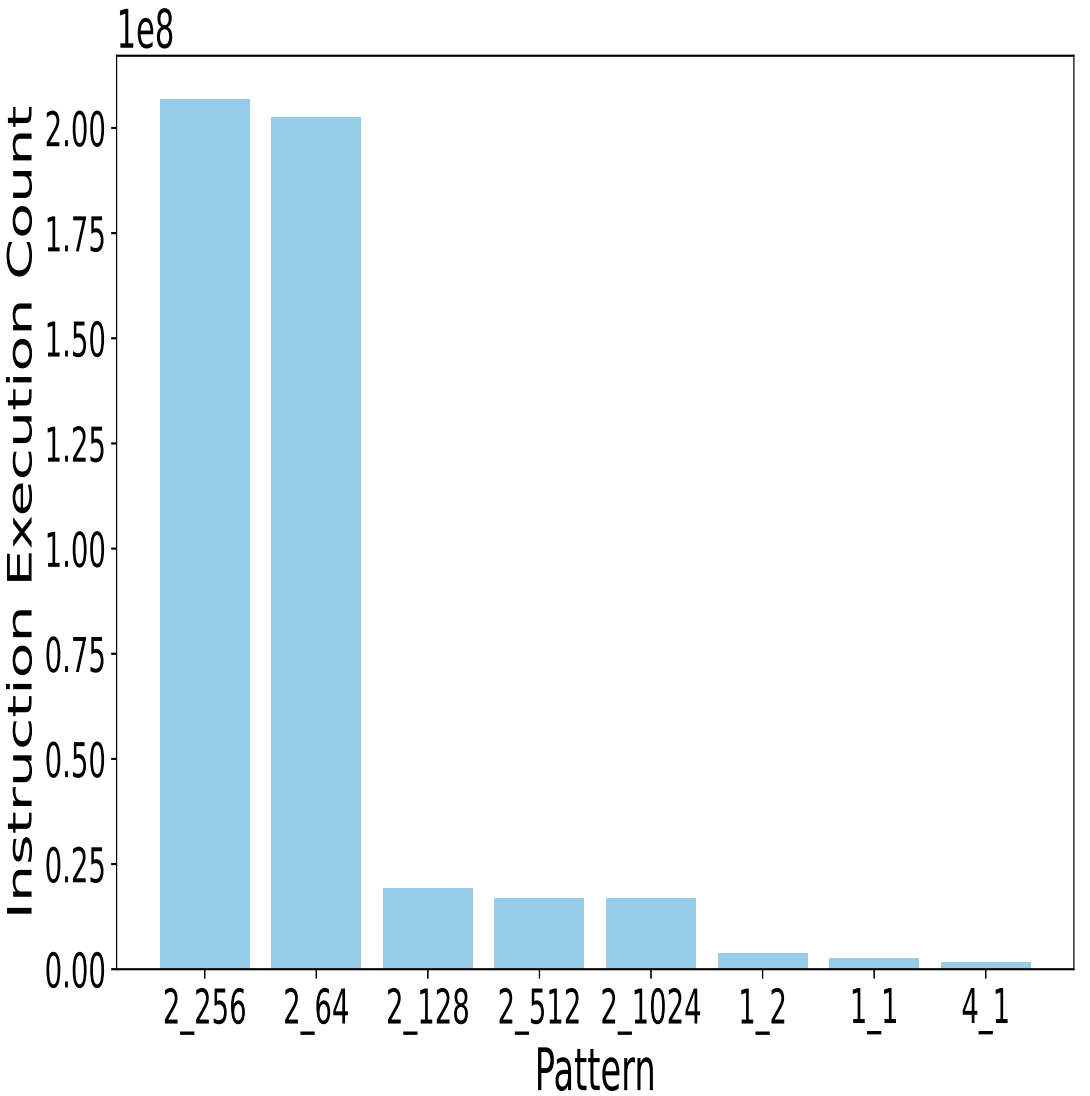}
    \caption*{\footnotesize{DenseNet121}}
  \end{minipage}
  \caption{Instruction execution count for different consecutive \texttt{addi} immediate values. The x-axis represents the pattern \texttt{X\_Y}, where \texttt{X} and \texttt{Y} denote the first and second immediate values in two consecutive \texttt{addi} instructions, respectively.}
  \label{cycle_count_consecutive_addi_mNV1}
  \vspace{-6mm}
\end{figure*}

This instruction merges the execution of two consecutive "add immediate" operations targeting different registers. However, incorporating two destination registers into the instruction syntax leaves only 15 bits available for encoding the immediate values. A direct split, allocating 7 bits per immediate (assuming signed values), would result in a limited range of (-64, 63), reducing potential use cases. To determine an optimal bit allocation, we analyzed the usage patterns of the \verb|addi| instruction within the inner convolution loops of the DNN models where this fused instruction would be most beneficial. As shown in Fig. \ref{cycle_count_consecutive_addi_mNV1}, all observed patterns mostly utilize only unsigned immediate values. Moreover, the most frequent patterns involve a small immediate followed by a larger one. Based on this observation, we allocated 5 bits to one immediate (i1: 0–31) and 10 bits to the other (i2: 0–1023), covering 100\% (for LeNet-5*), 86.03\% (for MobileNetV1), 75.19\% (for ResNet50), 66.89\% (for VGG16), 71.39\% (for MobileNetV2) and 95.13\% (for DenseNet121) of the relevant use cases when weighted by cycle count. This extension uses the CUSTOM 1 opcode. Listing~\ref{add2i_listing} shows the assembly syntax for this instruction, and Table~\ref{add2i_inst_decoding} shows its decoding scheme.
\vspace{-4mm}
\begin{figure}[h!]
\centering
\begin{minipage}{0.94\linewidth}
\begin{lstlisting}[caption={Assembly format of custom add2i instruction}, label={add2i_listing}, basicstyle=\small]
    ;rs1 = rs1 + i1
    ;rs2 = rs2 + i2
    add2i rs1, rs2, i1, i2
\end{lstlisting}
\end{minipage}
\vspace{-6mm}
\end{figure}

\begin{table}[h!]
    \centering
    \caption{Instruction decoding: add2i}
    \includegraphics[width=1\linewidth]{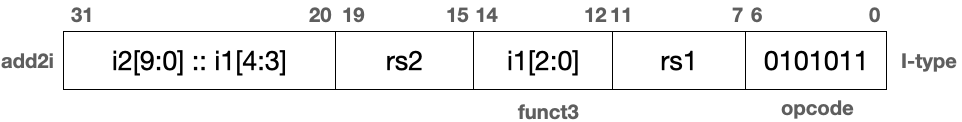}
    \label{add2i_inst_decoding}
    \vspace{-6mm}
\end{table}

\vspace{-6mm}
\subsubsection{fusedmac}
This custom instruction integrates the functionality of the previous two instructions. While it may seem having this instruction is redundant, it is necessary to evaluate whether using this combined instruction in all instances where either of the individual instructions could be applied is advantageous. Specifically, this instruction utilizes two functional units (\verb|mac| and \verb|add2i|), leading to increased power usage. Therefore, in scenarios where only the \verb|add2i| operation is needed, executing an unnecessary \verb|mac| operation on fixed registers will result in wasted power. Furthermore, an analysis of the baseline code reveals multiple opportunities to utilize all these custom extensions independently. As shown in the third subfigure of Fig.~\ref{pattern_count_plots}, the \texttt{fusedmac} pattern - comprising of four instructions in the sequence \texttt{addi}, \texttt{addi}, \texttt{mul}, and \texttt{add} - constitutes nearly 10\% of the total code. This substantial proportion suggests that this pattern recurs frequently enough to justify its own custom instruction and dedicated hardware acceleration. Moreover, the occurrence of this pattern increases with larger neural network models, where having a dedicated ISA extension like the \texttt{fusedmac} is justified.
Therefore, despite apparent redundancy, including all custom instructions is justified to optimize power efficiency and flexibility. This extension uses the CUSTOM 0 opcode. Listing~\ref{fusedmac_listing} shows the assembly syntax for this instruction and Table~\ref{fusedmac_inst_decoding} shows its decoding scheme.
\vspace{-4mm}
\begin{figure}[H]
\centering
\begin{minipage}{\linewidth}
\begin{lstlisting}[caption={Assembly format of custom fusedmac instruction}, label={fusedmac_listing}, basicstyle=\small]
    ;x20 = x20 + x21*x22
    ;rs1 = rs1 + i1
    ;rs2 = rs2 + i2
    fusedmac rs1, rs2, i1, i2
\end{lstlisting}
\end{minipage}
\vspace{-4mm}
\end{figure}
\begin{table}[h!]
    \centering
    \caption{Instruction decoding: fusedmac}
    \includegraphics[width=1\linewidth]{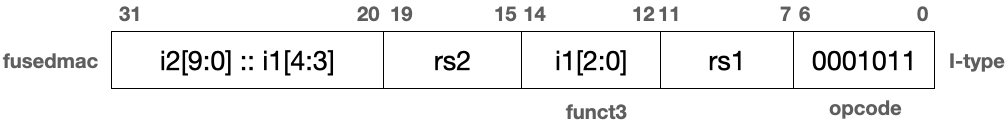}
    \label{fusedmac_inst_decoding}
    \vspace{-10mm}
\end{table}
\subsubsection{zol}
Profiling of C codes generated using the proposed flow on baseline RISC-V shows that the number of times \verb|blt| instruction is executed is 923.2K in LeNet-5*, 23.6M in MobileNetV1, 1.89B in ResNet50, 6.88B in VGG16, 153.76M in MobileNetV2, and 1.43B in DenseNet121, highlighting the need for its optimization. Because of the way TVM generates code, lengths of convolutional \verb|for| loops are known at compile time. Consequently, these loops can be fully unrolled, enabling the use of ISA extensions related to hardware loops to reduce cycle count. A hardware loop instruction essentially indicates in advance that the section of code following it is part of a loop, specifying the start and end addresses along with the iteration count. By leveraging hardware loops, the \verb|blt| instruction is entirely eliminated from loop execution, along with the explicit increment of the loop counter, which is instead managed by the hardware itself. In this processor version implementation, the instructions \verb|dlp|, \verb|dlpi|, \verb|zlp|, \verb|set.zc|, \verb|set.zs| and \verb|set.ze| are introduced, based on Synopsys' reference processor design. This requires modifications to the Program Control Unit (PCU) and the addition of three new registers: \verb|ZC| (Zero-overhead loop count), \verb|ZS| (Start address), and \verb|ZE| (End address). These registers are set up in the execution stage of the CPU pipeline. A total of five instructions are implemented to enable zero-overhead looping, with their decoding illustrated in Table~\ref{zol_inst_decoding}. The hardware loop extensions utilize two opcodes: 11101, reserved for hardware loops, and 10111, which remains available if the implementation does not require instructions exceeding 32 bits. 

\begin{figure*}[ht!]
  \centering
  \begin{minipage}{0.33\textwidth}
    \centering
    \includegraphics[width=\linewidth, height=3cm]{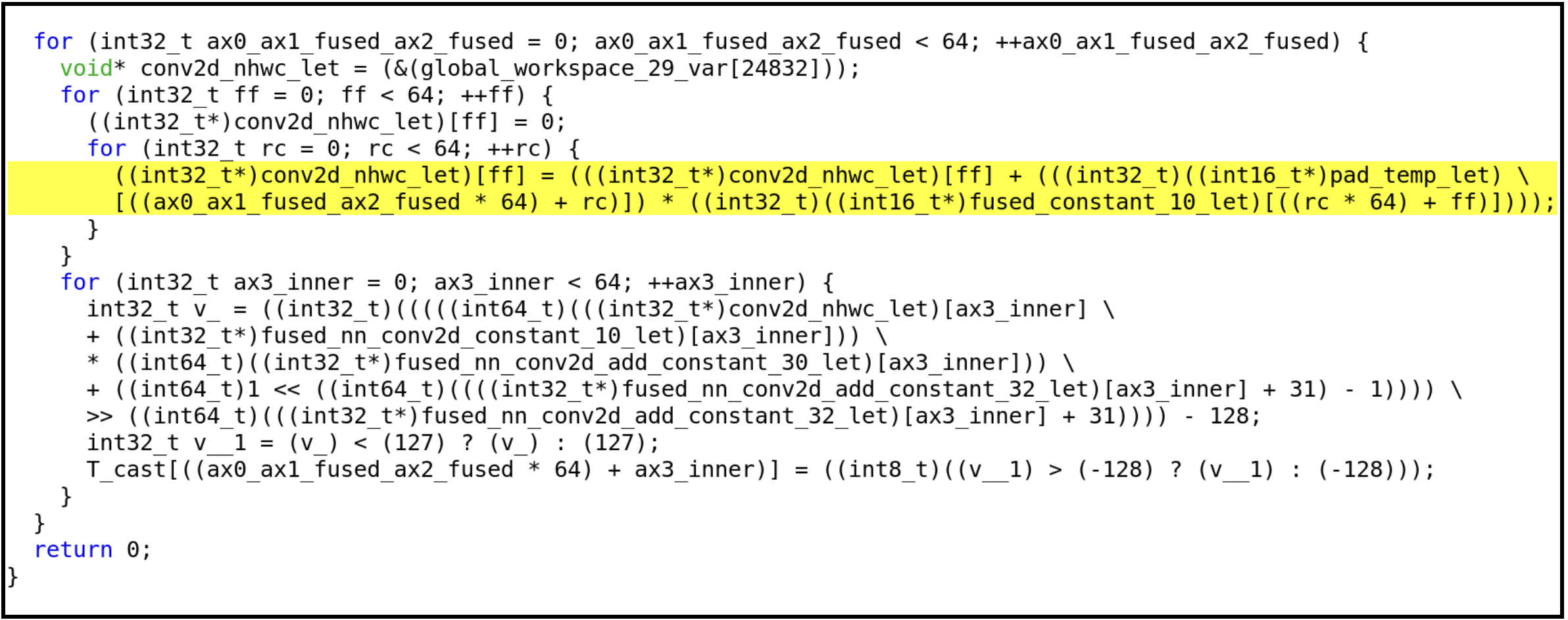}
    \caption*{\footnotesize{(a)}}
  \end{minipage}%
  \begin{minipage}{0.33\textwidth}
    \centering
    \includegraphics[width=\linewidth, height=3cm]{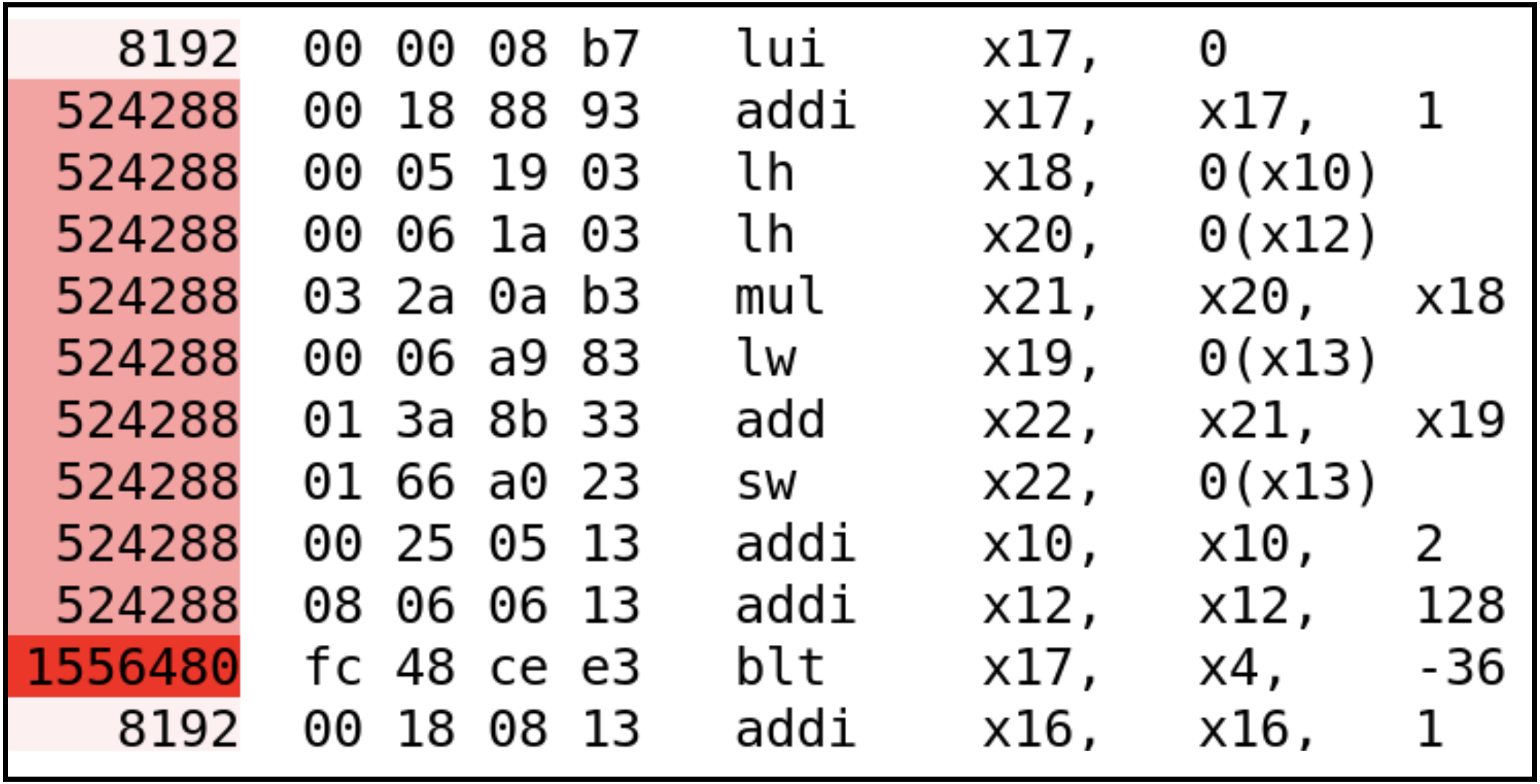}
    \caption*{\footnotesize{(b)}}
  \end{minipage}%
  \begin{minipage}{0.33\textwidth}
    \centering
    \includegraphics[width=\linewidth, height=3cm]{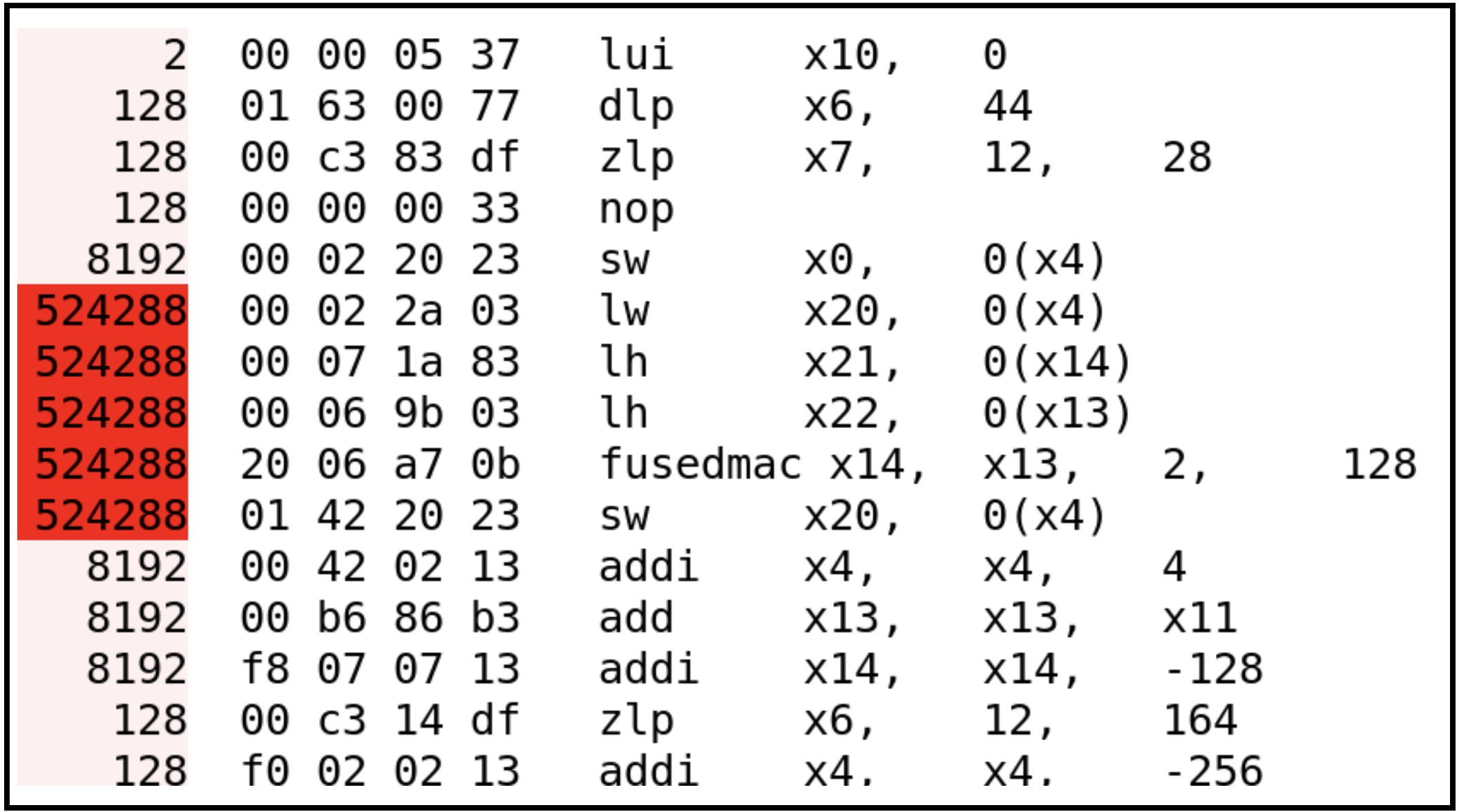}
    \caption*{\footnotesize{(c)}}
  \end{minipage}
  \caption{Flow-generated C code of a convolution operation (a) in MobileNetV1 alongside its corresponding assembly in the baseline processor (v0) (b) and the fully extended processor (v4) (c). \texttt{blt} instruction in baseline is eliminated as a result of the ISA extension targeting hardware loops (\texttt{zol})}
  \label{flow_C_to_asm_zol}
  \vspace{-6mm}
\end{figure*}

Fig. \ref{flow_C_to_asm_zol} presents a comparison of the assembly code generated for the highlighted C code, comparing processor versions v0 and v4. The highlighted columns in the last two figures indicate the total number of clock cycles spent executing the corresponding instructions in the test case. Notably, the removal of the \verb|blt| instruction, enabled by the hardware loop extension, leads to a significant reduction in execution cycles, contributing to improved efficiency.
\vspace{-3mm}
\begin{table}[h!]
    \centering
    \caption{Instruction decoding: zol}
    \includegraphics[width=0.9\linewidth, height=5cm]{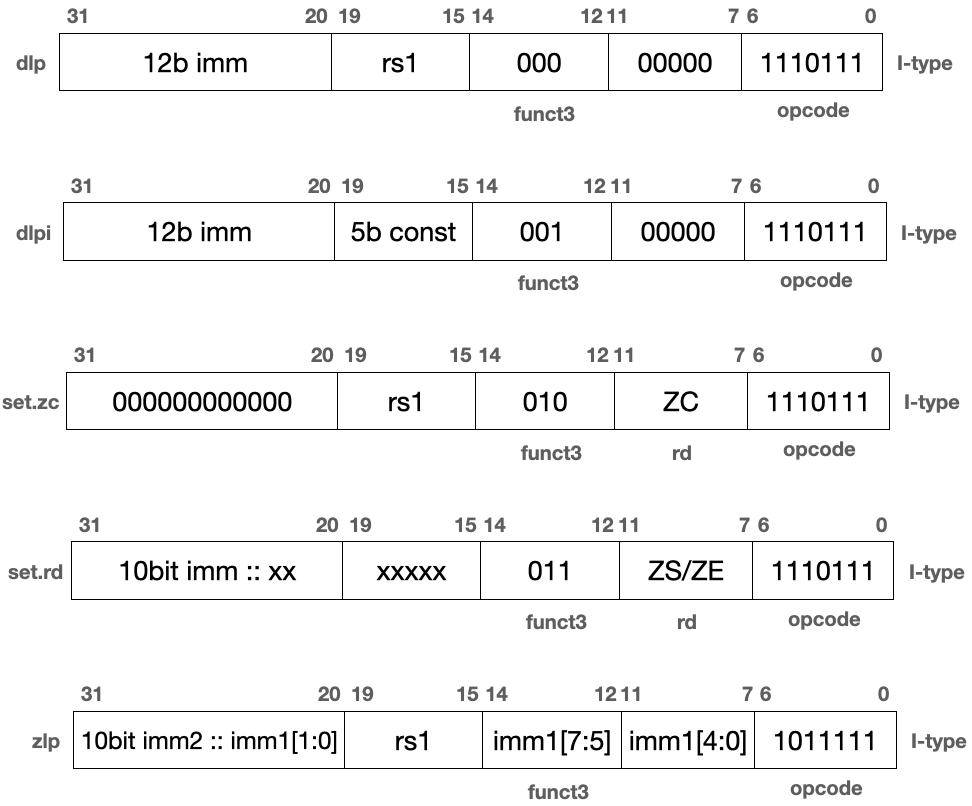}
    \label{zol_inst_decoding}
\end{table}

\subsection{Compilation}
To ensure the successful compilation of the generated C code using ASIP Designer, the project file must be configured appropriately. This requires an understanding of the Synopsys Chess\textsuperscript{\textregistered} compiler~\cite{chess_compiler}, which consists of two key components:

\begin{enumerate}
    \item \textbf{Front-End} – Responsible for parsing and interpreting the intent of the C source code.
    \item \textbf{Back-End} – Responsible for translating the parsed code into the corresponding assembly instructions for the target architecture.
\end{enumerate}

While the back-end of the Chess compiler is fixed, users can choose between two available front-ends:
\begin{itemize}
    \item LLVM Front-End
    \item Chess Front-End
\end{itemize}

Additionally, ASIP Designer allows different front-ends to be used for different source files within the same project. The selection of the appropriate front-end depends on specific characteristics of the generated C code, as outlined below:

\begin{itemize}
    \item Handling the \verb|__attribute__| Construct: The generated C code makes use of the \verb|__attribute__| construct, which is a standard feature in C but is not supported by the Chess front-end. However, it is fully supported by the LLVM front-end. Therefore, any source file that contains the \verb|__attribute__| keyword must be compiled using LLVM.
    \vspace{1mm}
    \item Optimizing Custom Instruction Utilization: A key drawback of the LLVM front-end is its inability to automatically identify and replace groups of assembly instructions with custom processor instructions. In contrast, the Chess front-end can efficiently infer and apply custom instructions for performance optimizations. As a result, source files that could benefit from custom instruction optimizations should be compiled using the Chess front-end.
\end{itemize}

The Synopsys Chess compiler can be retargeted to a specific processor by defining ``\verb|chess_rewrite|'' rules which specify how the custom instructions should be used.  It should be noted that the actual PDG implementations of these instructions are opaque to the Chess compiler. The ``\verb|chess_rewrite|'' rule applied to the ``\verb|mac|'' instruction is as shown in listing \ref{chess_rewrite_mac_listing}:
\begin{lstlisting}[numbers=none, deletekeywords={int}, caption={Chess rewrite rule for mac instruction}, label={chess_rewrite_mac_listing}, basicstyle=\small]
chess_rewrite int mac_rule(int c, int a, int b) 
{return c + a*b;} -> {return MAC(c,a,b);}			  
\end{lstlisting}
\par These rules allow the Chess compiler to identify patterns where the use of custom instructions may be more efficient than the baseline ISA instructions. We configured the Chess compiler so that a fully optimized C application is generated. This ensured compatibility with embedded RISC-V processors while leveraging the custom instructions of the target processor architecture for enhanced performance.

\subsection{Hardware Implementation}
\subsubsection{RTL generation from ASIP Designer}
\begin{figure}
    \centering
    \includegraphics[width=\linewidth]{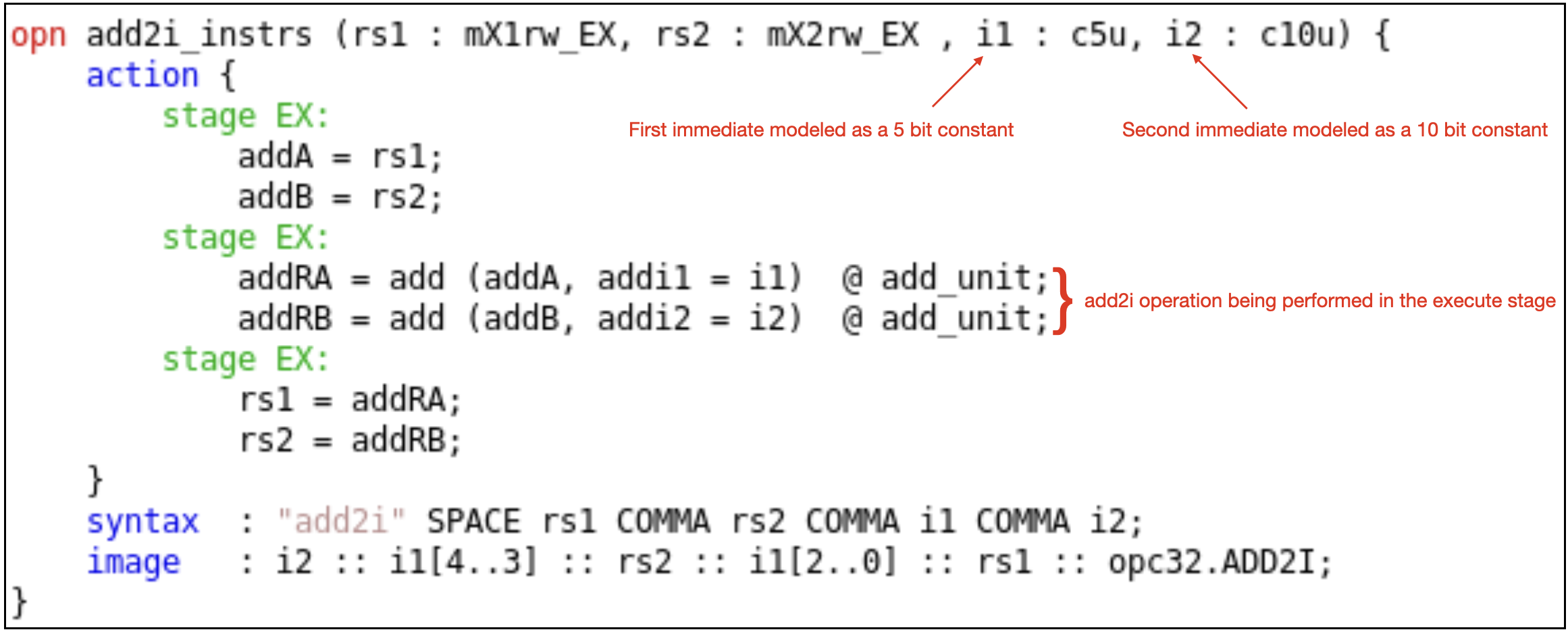}
    \caption{nML model of \texttt{fusedmac} extension}
    \label{fusedmac_nML}
    \vspace{-6mm}
\end{figure}

\begin{figure}[H]
  \centering
  \includegraphics[width=0.47\textwidth]{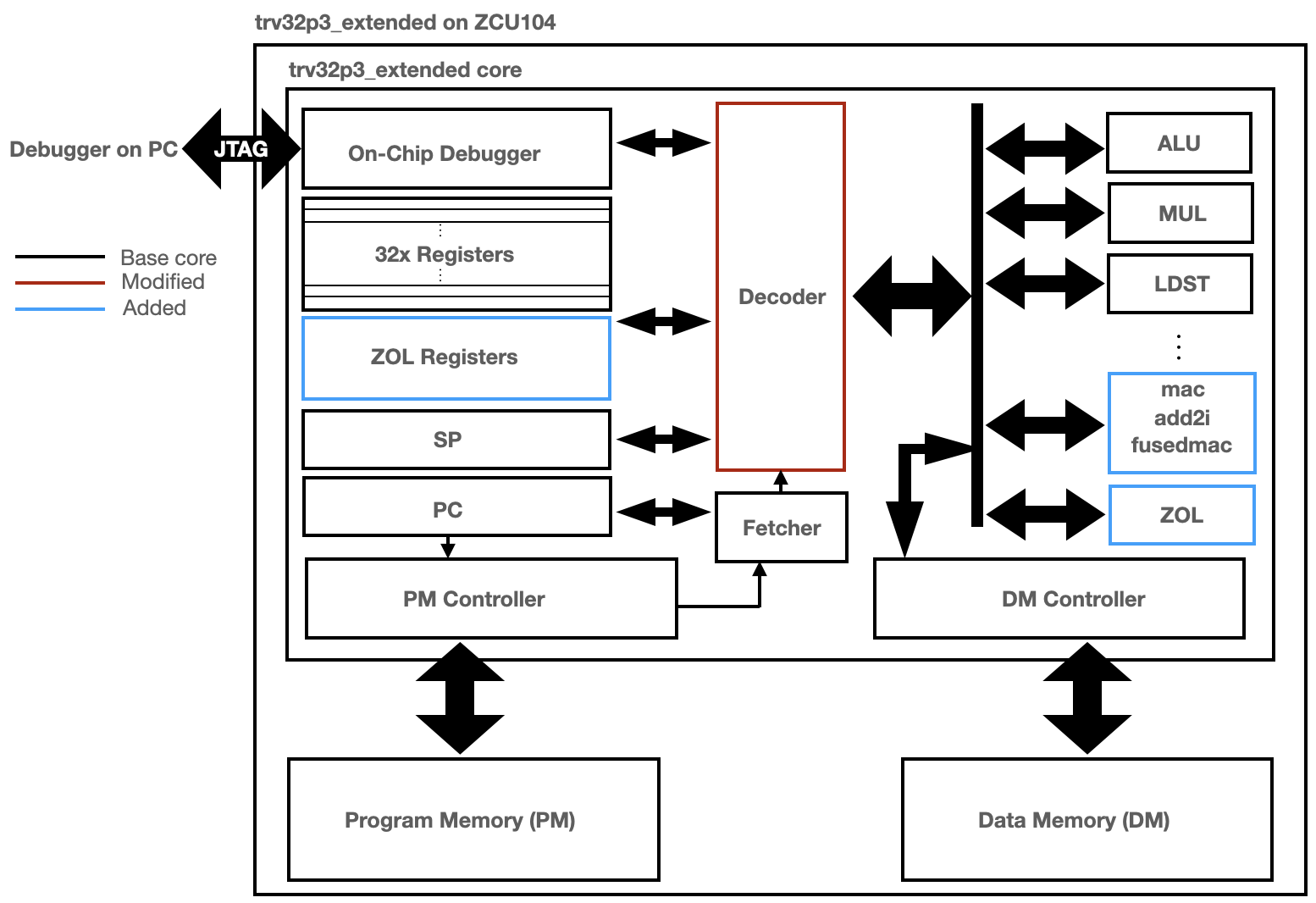}
  \caption{\label{trv32p3_ext} Custom RISC-V illustrating modifications on the baseline}
\end{figure}

\begin{figure*}[t]
    \centering
    \begin{subfigure}[t]{0.32\textwidth}
        \centering
        \includegraphics[height=3.5cm, width=\linewidth]{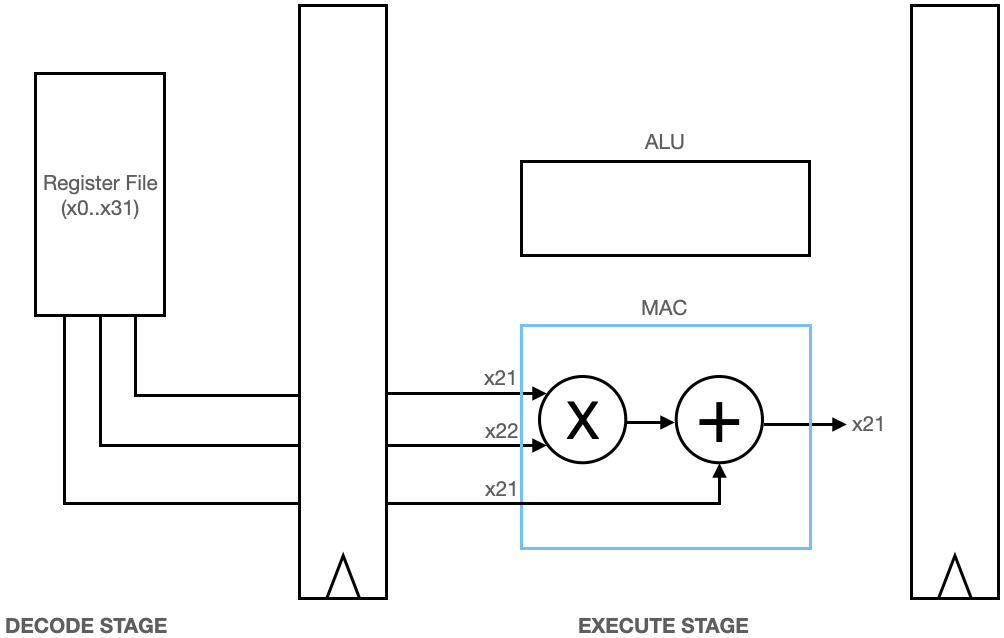}
        \caption*{(a) \texttt{mac} hardware generated from nML}
    \end{subfigure}
    \hfill
    \begin{subfigure}[t]{0.32\textwidth}
        \centering
        \includegraphics[height=3.5cm, width=\linewidth]{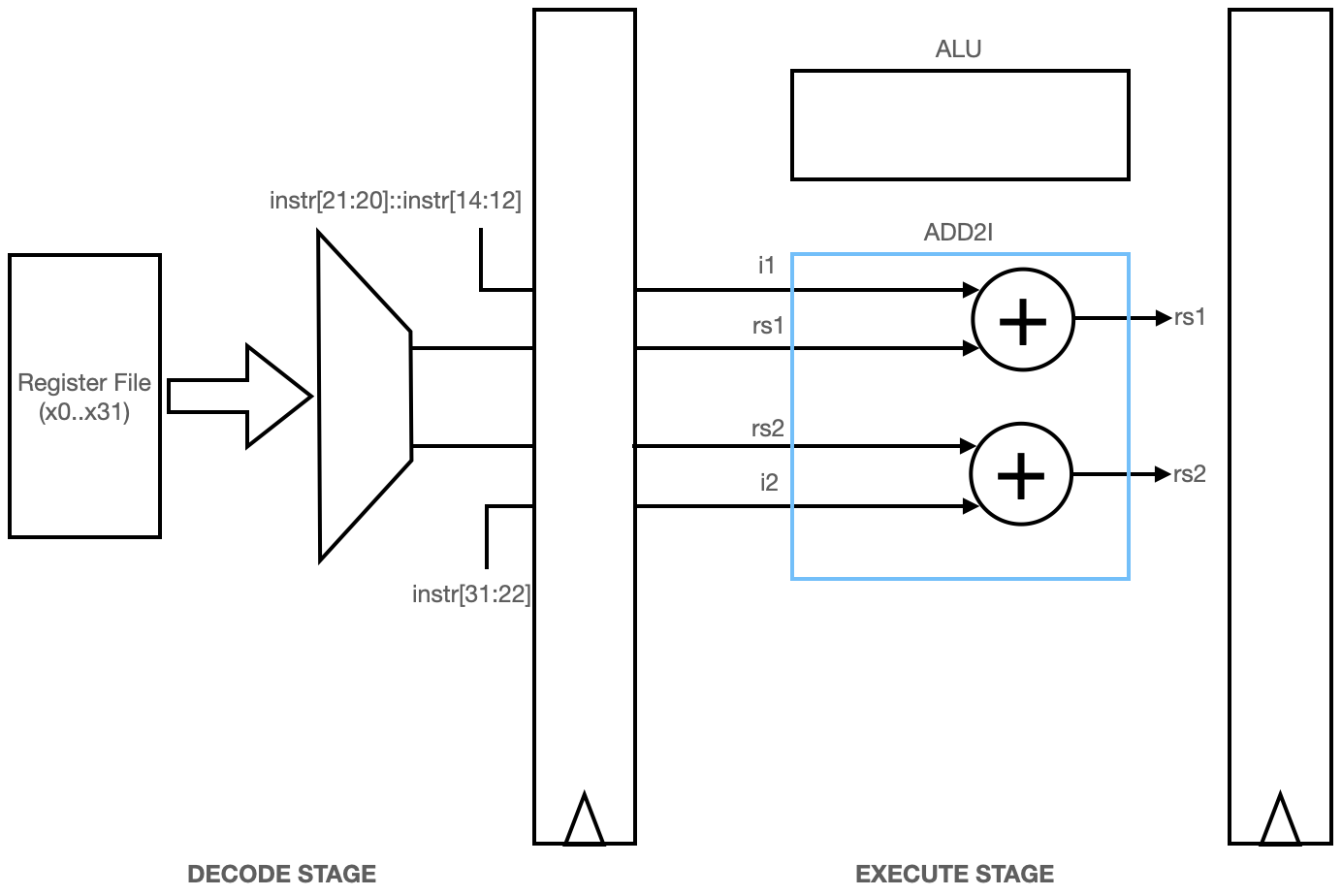}
        \caption*{(b) \texttt{add2i} hardware generated from nML}
    \end{subfigure}
    \hfill
    \begin{subfigure}[t]{0.32\textwidth}
        \centering
        \includegraphics[height=3.5cm, width=\linewidth]{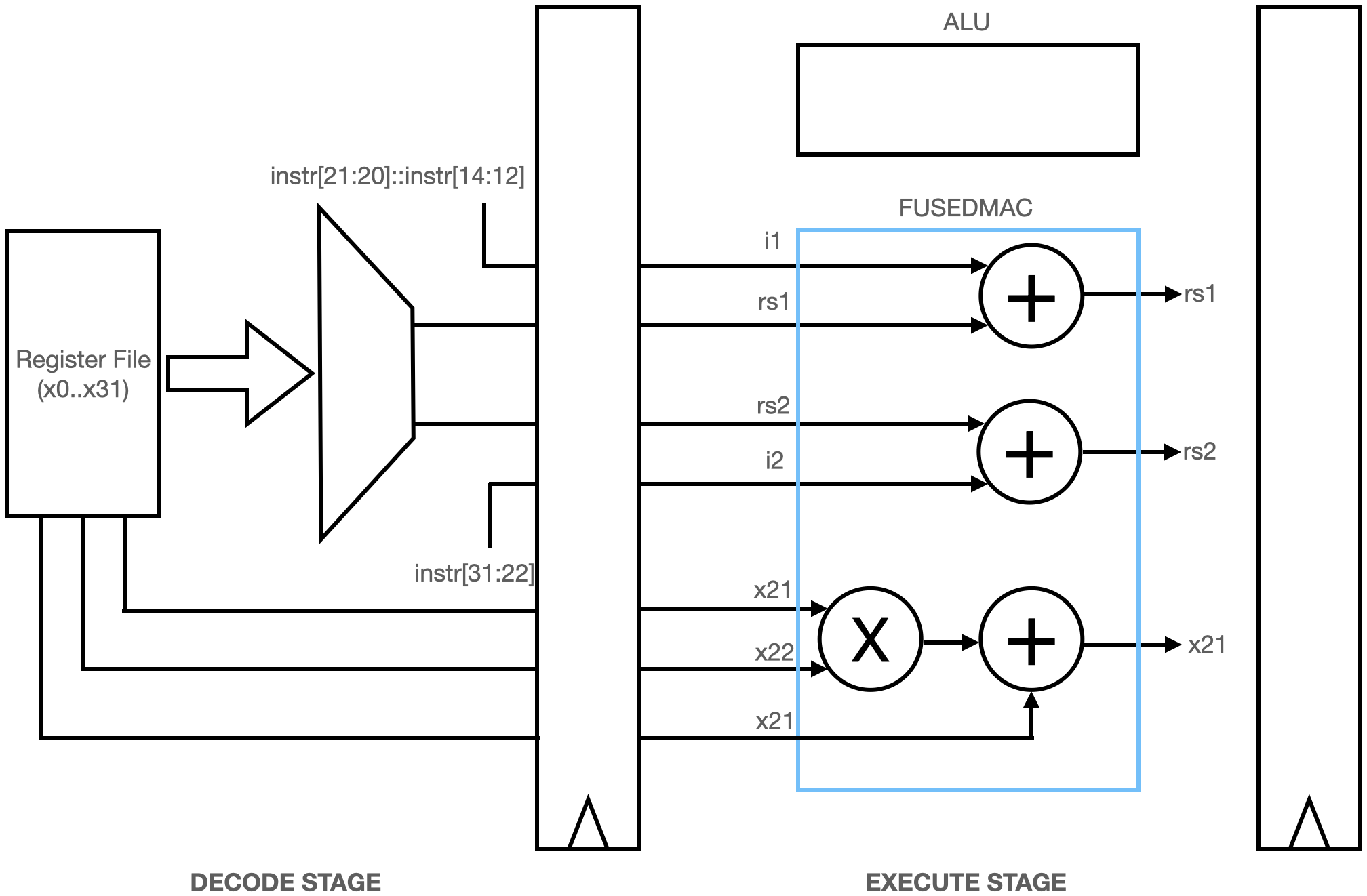}
        \caption*{(c) \texttt{fusedmac} hardware generated from nML}
    \end{subfigure}
    \caption{Hardware diagrams generated from nML for \texttt{fusedmac}, \texttt{mac}, and \texttt{add2i} instructions.}
    \label{nML_to_HW}
\end{figure*}

The extensions obtained from Step 2 (in Fig.~\ref{top_level_flow_block_diagram}) of the flow are modeled in nML, with Fig.~\ref{fusedmac_nML} presenting snippet of the \texttt{fusedmac} extension. Using the Go compiler, a synthesizable Verilog representation of the extended core is generated from this nML modeling of the obtained extensions. The corresponding hardware implementation for the \texttt{mac}, \texttt{add2i} and \texttt{fusedmac} extensions is depicted in Fig.~\ref{nML_to_HW}. Additionally, for debugging purposes, On-Chip Debug (OCD) support is enabled by configuring the Go compiler to incorporate a JTAG interface within the processor module. Fig.~\ref{trv32p3_ext} shows the key functional units within the extended trv32p3 processor.
\par The trv32p3 core implements a modified Harvard architecture.  The memory interfaces that provide access to the data memory and program memory are defined in PDG code. We implement RAM blocks independent of the processor module that are compatible with the interfaces defined in the PDG code. In this design, the processor data memory and program memory are implemented with the dedicated block RAM available within the ZCU104 FPGA.  Logic specific to the target hardware must be added to provide an interconnection between the block RAM and the trv32p3 memory interfaces.  In this case, the block RAM output register must be disabled.  This reduces the read operation latency to a single clock cycle, as required by the trv32p3 interface.

\vspace{-10mm}
\subsubsection{Implementation in Vivado}
The hardware description generated by ASIP Designer is synthesized and implemented in Vivado (version 2024.2), as illustrated in Step 3 of Fig.~\ref{top_level_flow_block_diagram}. During this phase, a top-level testbench is provided, defining the clock, reset, and JTAG interface signals. Additionally, synthesis and implementation constraints are specified to map the design pins to the corresponding physical pins on the ZCU104 FPGA.
\vspace{-6mm}
\subsubsection{On-Chip Debugging} 
\begin{figure}
    \centering
    \includegraphics[width=\linewidth]{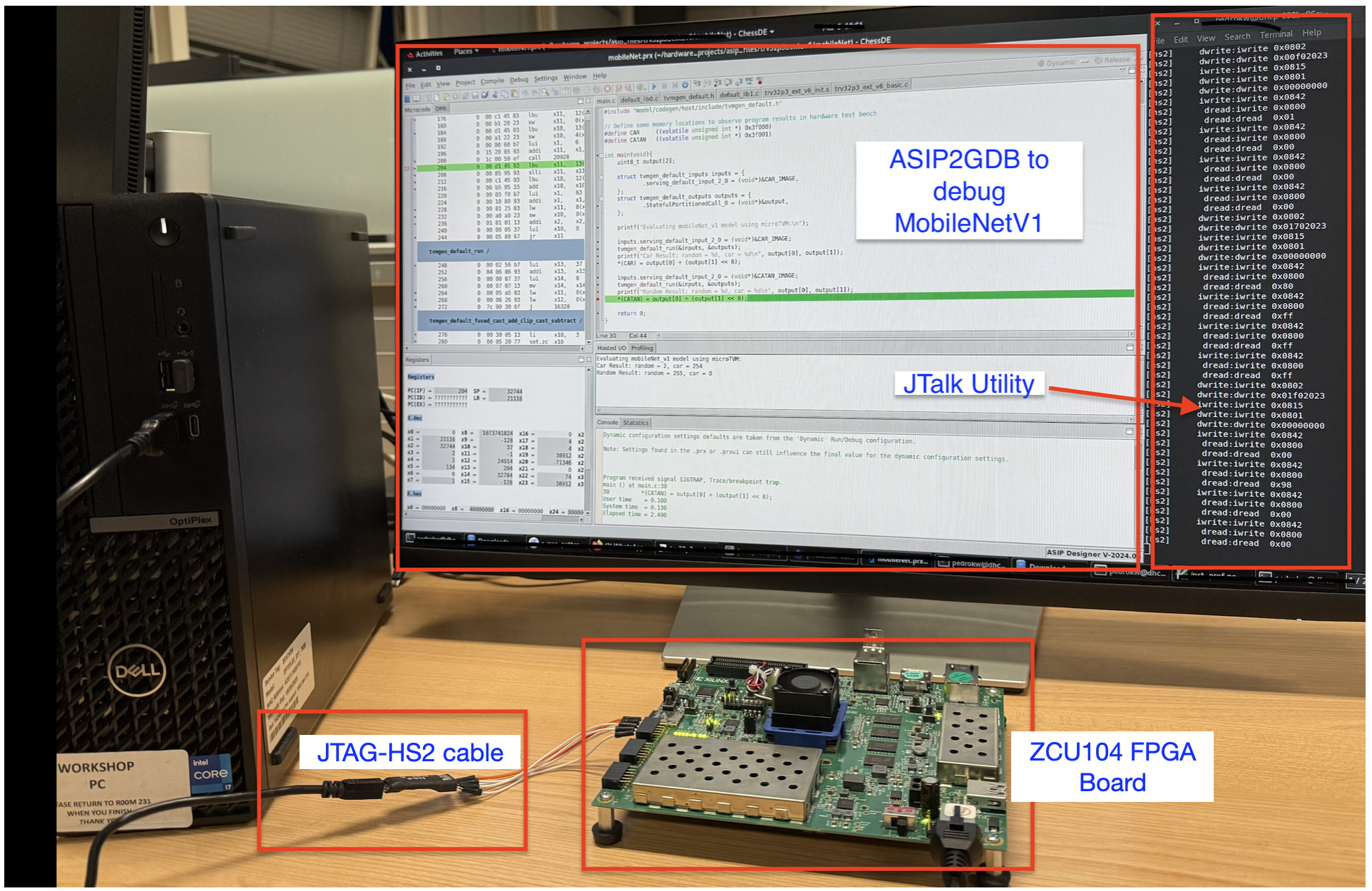}
    \caption{Experimental setup showing MobileNetV1 deployed and debugged on RISC-V through JTAG}
    \label{fpga_setup}
    \vspace{-4mm}
\end{figure}

The JTalk\textsuperscript{\textregistered} cable driver is used to facilitate the on-chip debugging of applications running on the target hardware (step 4 in Fig.~\ref{top_level_flow_block_diagram}). This work uses the Diligent\textsuperscript{\textregistered} JTAG HS2 cable. Synopsys ASIP2GDB debugging utility is used to communicate with the FPGA connected to the host PC running JTalk utility as shown in Fig.~\ref{fpga_setup}

\section{Results}
\par The performance improvements and resource utilisations associated with the extended RISC-V core discussed in Section (\ref{ISA_extensions}) are evaluated in this section. The performance is benchmarked with C implementations of CNN models obtained through the flow as described in Section (\ref{proposed_flow}).  The performance improvement and memory utilisation (shown in Table \ref{tab:memory_comparison}) associated with each of the five processor model variations are compared. 

\subsection{Implementation}
\begin{figure}
  \centering
  \begin{minipage}{0.2\textwidth}
    \centering
    \includegraphics[width=\linewidth]{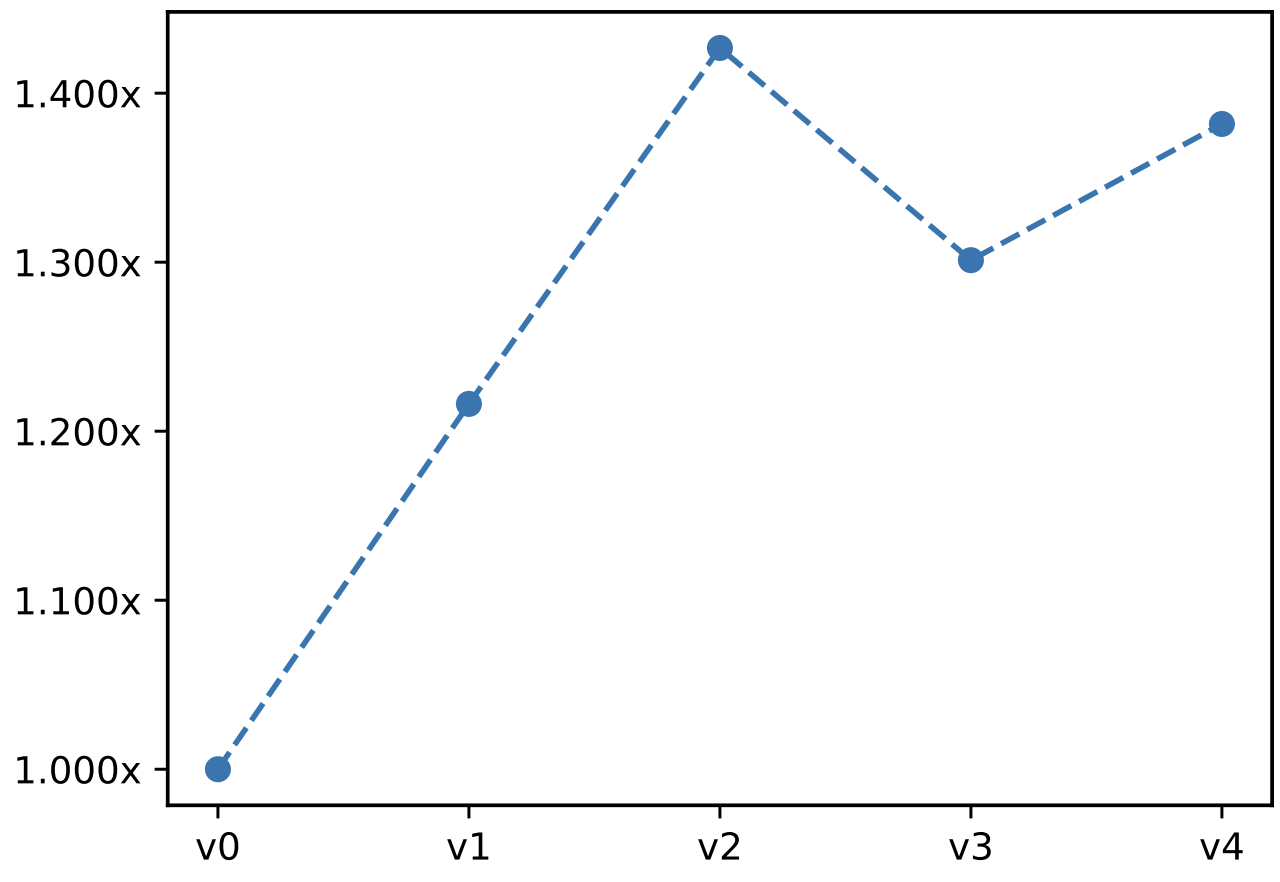}
    \caption*{\footnotesize{(a) Look-Up Tables}}
    \vspace{-1mm}
  \end{minipage}%
  \begin{minipage}{0.2\textwidth}
    \centering
    \includegraphics[width=\linewidth]{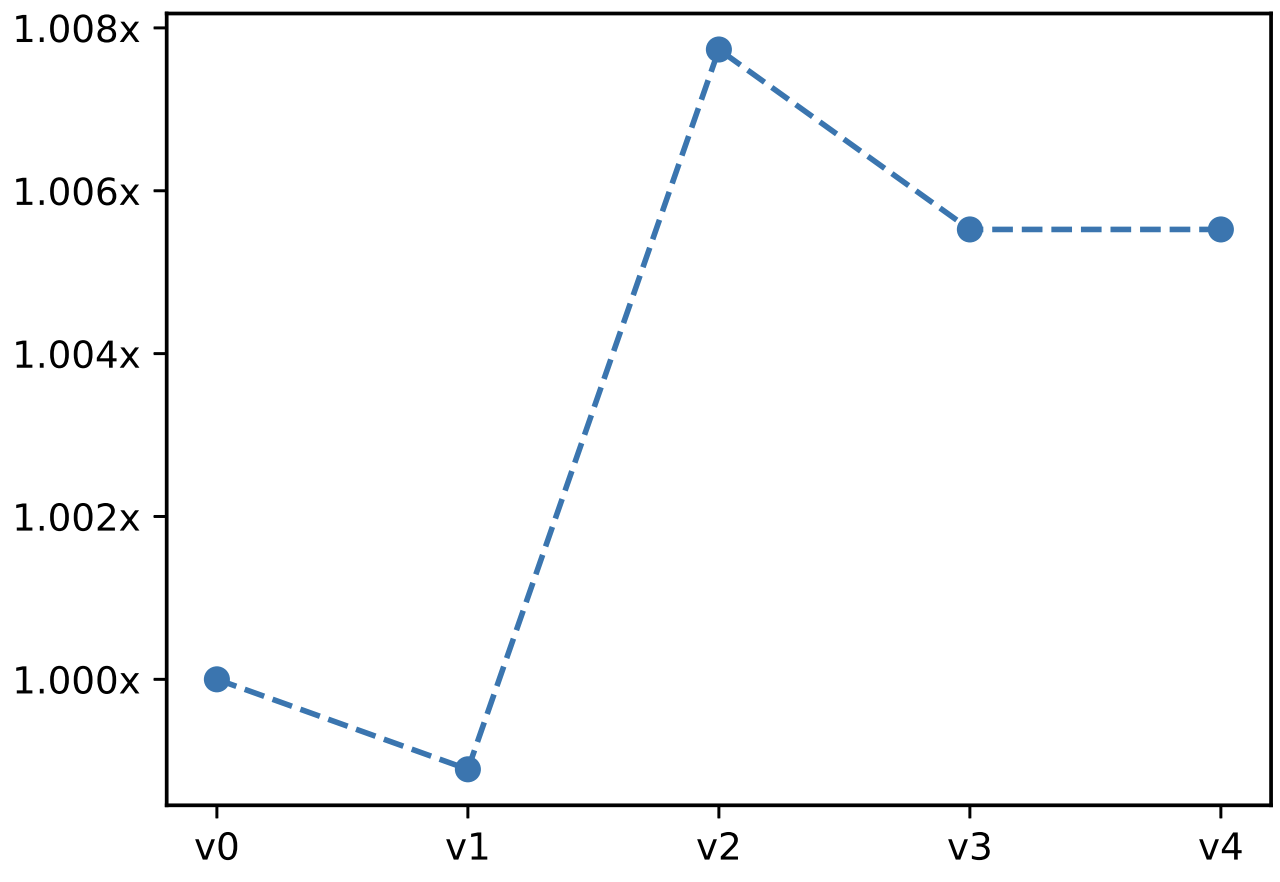}
    \caption*{\footnotesize{(b) Multiplexers}}
    \vspace{-1mm}
  \end{minipage}
  \begin{minipage}{0.2\textwidth}
    \centering
    \includegraphics[width=\linewidth]{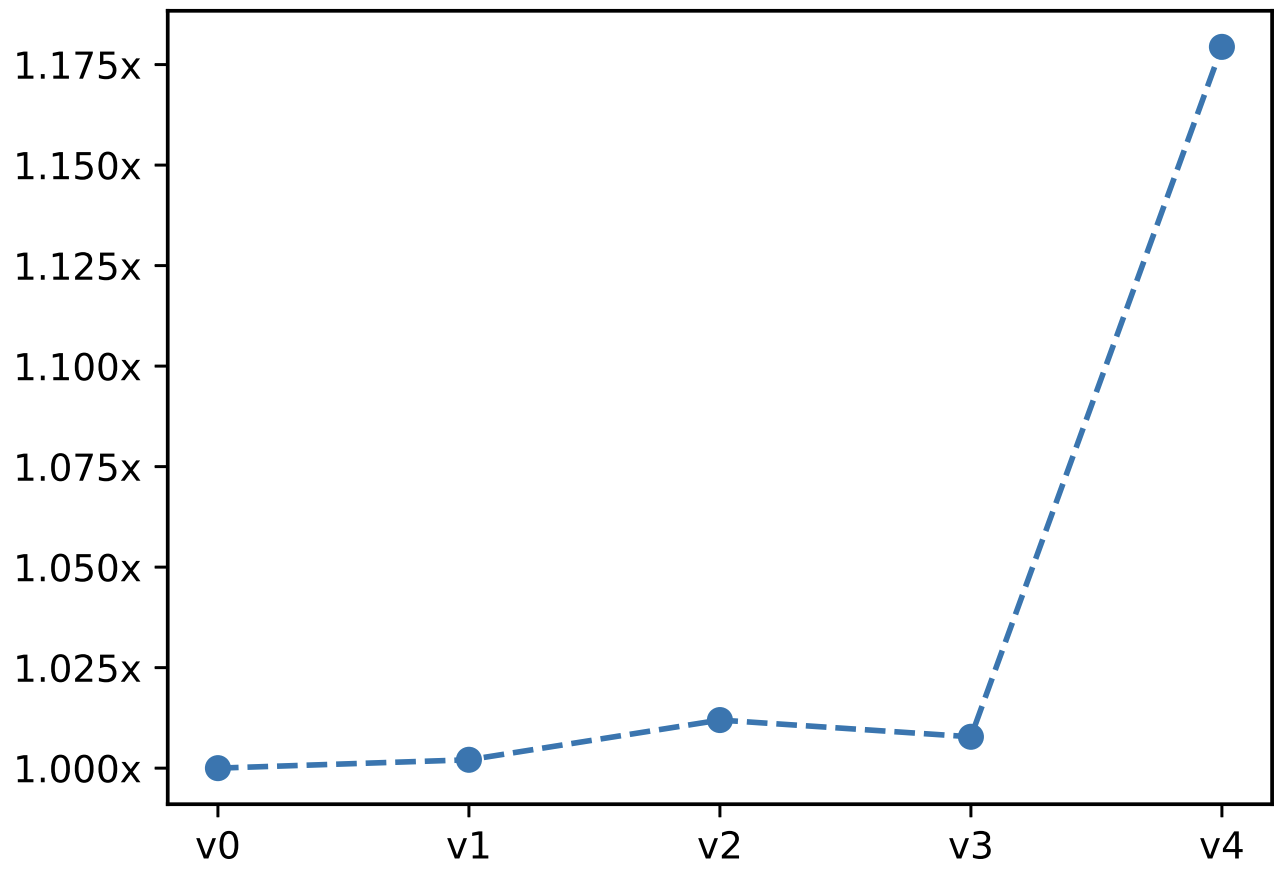}
    \caption*{\footnotesize{(c) Registers count}}
  \end{minipage}%
  \begin{minipage}{0.2\textwidth}
    \centering
    \includegraphics[width=\linewidth]{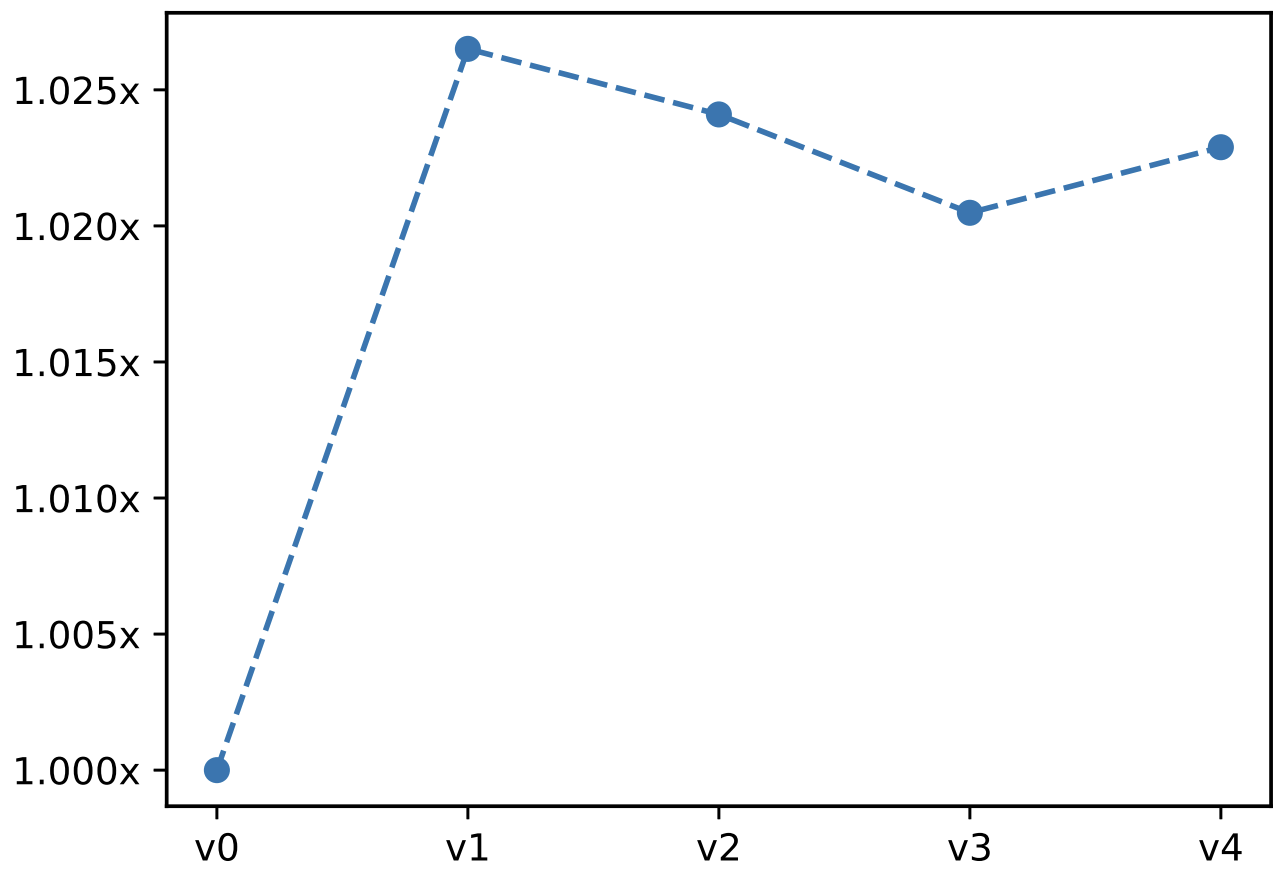}
    \caption*{\footnotesize{(d) Power}}
  \end{minipage}
  \caption{FPGA resource utilisation and power consumption with the addition of each successive extension, shown as a proportion of base core utilisation.}
  \label{util_graph}
  \vspace{-4mm}
\end{figure}

The resource utilisation associated with each processor model variation is shown in both Fig.~\ref{util_graph} and Table~\ref{resource_util}. The extended core (v4) comes with a LUT overhead of \textbf{38.17\%} when compared to the baseline core. However, an overhead of only \textbf{0.5\%} for multiplexers and \textbf{17.94\%} for registers is associated with the fully extended core. The majority of the overhead comes from the addition of the \verb|mac| and \verb|add2i| instructions, with the \verb|fusedmac| and \verb|zol| instructions being more resource efficient in terms of LUT usage.

\begin{table}
  \caption{FPGA utilisation of all processor variants}
  \label{resource_util}
  \scriptsize
  \begin{adjustbox}{max width=\linewidth}
  \begin{tabular}{lrrrrr}
    \toprule
    \textbf{Processor}&LUT&MUX&Registers&DSP&Power\\
    \midrule
    \textbf{v0: Baseline} & 4,492 & 905 & 1,923 & 4 & 830 mW\\
    \textbf{v1: v0 + mac} & 5,463 & 904 & 1,927 & 7 & 852 mW\\
    \textbf{v2: v1 + add2i} & 6,409 & 912 & 1,946 & 7 & 850 mW\\
    \textbf{v3: v2 + fusedmac} & 5,845 & 910 & 1,938 & 7 & 847 mW\\
    \textbf{v4: v3 + hardware loops} & 6,207 & 910 & 2,268 & 7 & 849 mW\\
    \midrule
  \textbf{Overhead:} & 1,715 & 5 & 345& 3 & 19 mW\\
  &  (38.17\%) & (0.5\%) & (17.94\%) & (75\%)& (2.28\%)\\
  \bottomrule
\end{tabular}
\end{adjustbox}
\end{table}
\par The power utilisation is estimated in the post-implemented design with typical process corner and default switching activity. The extended core (v4) is power efficient with an estimated power overhead of only \textbf{2.28\%}.

\subsection{Performance Evaluation}

Benchmarking was conducted on six neural network models: MobileNetV1, ResNet50, VGG16, MobileNetV2, and DenseNet121, generated using the proposed flow, along with a hand-coded C implementation of an LeNet-5-like CNN (denoted as LeNet-5* in this paper). The architecture of LeNet-5* model is detailed in Table \ref{mnistCnnModel}. The inclusion of a hand-coded model demonstrates that the proposed extensions can be applied to both tool-generated and manually written C code, ensuring independence from specific tool artifacts. As long as the required patterns exist within the provided C code, the extensions remain applicable.

\begin{figure}
    \centering
    \includegraphics[width=\linewidth]{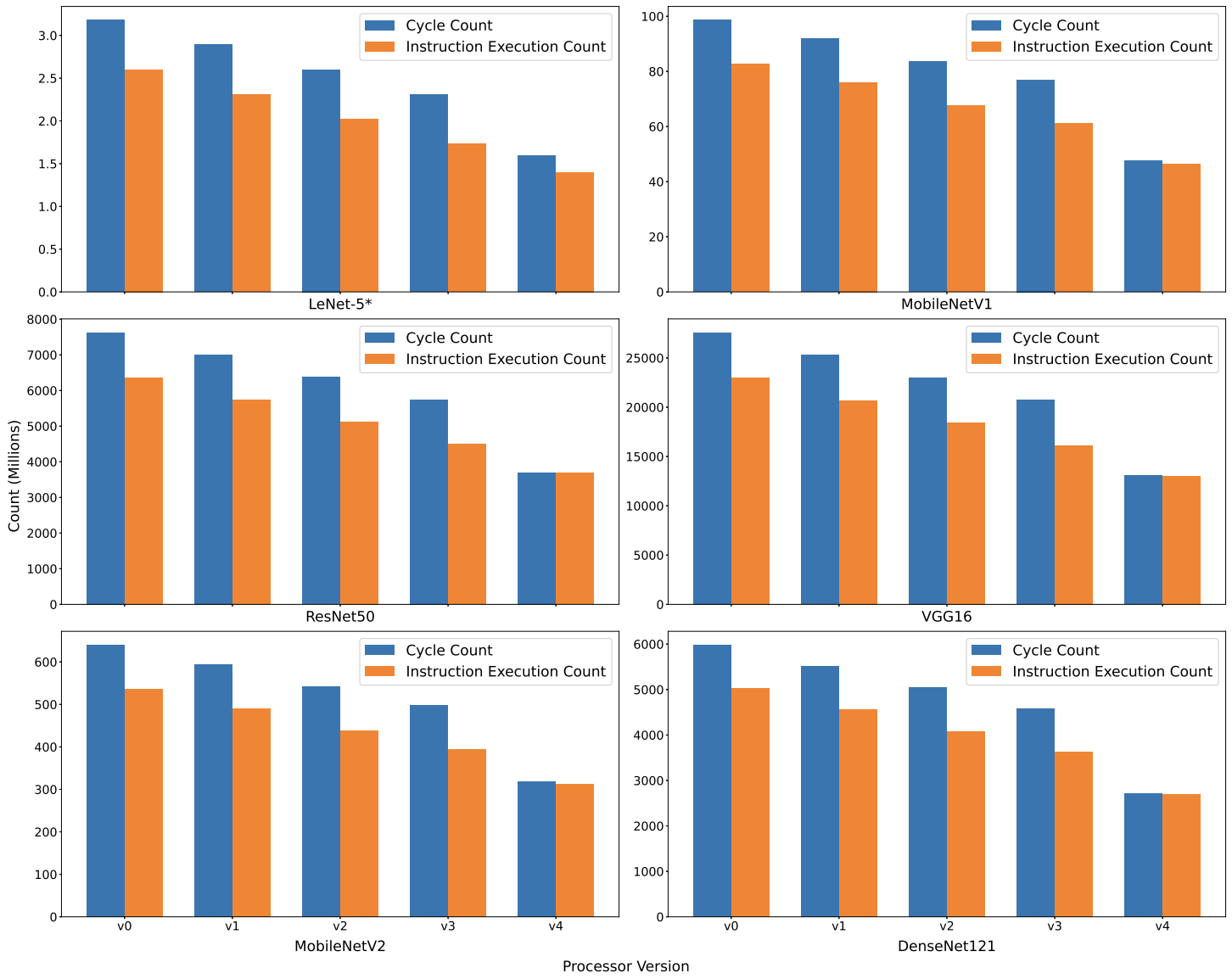}
    \caption{Cycle and Instruction count for inferences across DNN models on five RISC-V variants}
    \label{cycle_and_inst_count}
    \vspace{-6mm}
\end{figure}

\begin{table}
\vspace{-3mm}
\centering
\caption{\label{mnistCnnModel}LeNet-5* model structure}
\scriptsize
\begin{adjustbox}{max width=\linewidth}
\begin{tabular}{llll}
\toprule
& First Layer & Second Layer & Third Layer\\
\midrule
\textbf{Layer Type:} & Convolution & Convolution & MLP\\
\textbf{Input Size:} & 28$\times$28 & 12$\times$12$\times$12 & 32$\times$4$\times$4\\
\textbf{Output Size:} & 12$\times$12$\times$12 & 32$\times$4$\times$4 & 10$\times$1\\
\textbf{Activation:} & ReLU & ReLU & Softmax\\ 
\textbf{Kernel Size:} & 6$\times$6 & 6$\times$6 & N/A\\
\textbf{Stride:}      & 2 & 2 & N/A\\
\textbf{Num. Filters:} & 12 & 32 & N/A\\
  \bottomrule
\end{tabular}
\end{adjustbox}
\end{table}
Simulations were performed using ASIP Designer as well as a hardware testbench implemented in Xillinx Vivado.  The results from both simulation environments were identical and are presented in Fig.~\ref{cycle_and_inst_count}. The reported cycle count represents the average number of cycles required for two CNN inference operations.

\begin{table*}
    \centering
    \caption{Comparison of Data and Program Memory Usage Across Processor Versions}
    \label{tab:memory_comparison}
    \setlength{\tabcolsep}{3pt}
    \scriptsize
    \begin{tabular}{lcccccccccccc}
        \toprule
        \multirow{2}{*}{\textbf{Processor}} & \multicolumn{2}{c}{\textbf{LeNet-5*}} & \multicolumn{2}{c}{\textbf{MobileNetV1}} & \multicolumn{2}{c}{\textbf{ResNet50}} & \multicolumn{2}{c}{\textbf{VGG16}} & \multicolumn{2}{c}{\textbf{MobileNetV2}} & \multicolumn{2}{c}{\textbf{DenseNet121}} \\
        \cmidrule(lr){2-3} \cmidrule(lr){4-5} \cmidrule(lr){6-7} \cmidrule(lr){8-9} \cmidrule(lr){10-11} \cmidrule(lr){12-13}
        & \textbf{DM (kB)} & \textbf{PM (kB)} & \textbf{DM (kB)} & \textbf{PM (kB)} & \textbf{DM (MB)} & \textbf{PM (kB)} & \textbf{DM (MB)} & \textbf{PM (kB)} & \textbf{DM (MB)} & \textbf{PM (kB)} & \textbf{DM (MB)} & \textbf{PM (kB)} \\
        \midrule
        \textbf{v0: Baseline} & 60.83 & 1.47 & 622.04 & 24.19 & 43.65 & 39.50 & 15.76 & 13.18 & 4.19 & 40.86 & 14 & 105.14 \\
        \textbf{v1: v0 + mac} & 31.53 & 1.46 & 592.74 & 24.12 & 43.62 & 39.28 & 15.73 & 13.21 & 4.19 & 40.70  & 14 & 104.39 \\
        \textbf{v2: v1 + add2i} & 31.53 & 1.44 & 592.74 & 23.08 & 43.62 & 37.94 & 15.73 & 12.82 & 4.19 & 39.15 & 14 & 100.96 \\
        \textbf{v3: v2 + fusedmac} & 31.53 & 1.43 & 592.74 & 22.97 & 43.62 & 37.73 & 15.73 & 12.77 & 4.19 & 38.97 & 14 & 100.49 \\
        \textbf{v4: v3 + hardware loops} & 31.48 & 1.32 & 592.74 & 21.87 & 43.62 & 35.50 & 15.73 & 12.31 & 4.19 & 36.72 & 14 & 102.51 \\
        \midrule
        \textbf{Total Memory Saved (\%)} & 48.24 & 10.20 & 4.71 & 9.59 & 0.06 & 10.12 & 0.19 & 6.60 & 0 & 10.13 & 0 & 2.50 \\
        \bottomrule
    \end{tabular}
\end{table*}

\begin{table*}[h]
    \centering
    \scriptsize
    \caption{Comparison of Related Works on ISA Extensions and HW-SW Co-Design for Edge AI}
    \label{related_works_table}
    \begin{threeparttable}  %
        \begin{tabular}{@{}lccccc@{}}  %
            \toprule
            \textbf{Work} & \textbf{Input Format} & \textbf{Modifications to Input} & \textbf{ISA Extensions} & \textbf{HW-SW Co-Design}  & \textbf{Automated End-to-End Flow} \\ 
            \midrule
            BARVINN \cite{10.1145/3566097.3567872} & ONNX & Required\footnotemark[1] & Generic & \multicolumn{1}{c}{No} & \multicolumn{1}{c}{No}  \\ 
            FlexACC \cite{9516466} & C & Required\footnotemark[2] & Model class-specific & \multicolumn{1}{c}{No} & \multicolumn{1}{c}{No} \\ 
            RISC-VTF \cite{9658643} & Assembly & N/A & Transformer-specific & \multicolumn{1}{c}{No} & \multicolumn{1}{c}{No} \\ 
            XPulpNN \cite{xpulp} & C\footnotemark[3] & Required  & Generic & \multicolumn{1}{c}{No} & \multicolumn{1}{c}{No} \\ 
            PULP \cite{7864441} & C & Required & DSP-specific & \multicolumn{1}{c}{No} & \multicolumn{1}{c}{No} \\ 
            AI-RISC\cite{verma2022ai} & DSL & Required & Generic & Modifies TVM-based flow & \multicolumn{1}{c}{Yes} \\ 
            LiteAIR5\cite{10257058} & DSL & Required & Generic & Modifies TVM-based flow & \multicolumn{1}{c}{Yes} \\ 
            \textbf{Ours} & DSL & \multicolumn{1}{c}{No} & Model class-specific & \multicolumn{1}{c}{Yes} & \multicolumn{1}{c}{Yes} \\ 
            \bottomrule
        \end{tabular}
   \vspace{2mm} %
    \noindent\textsuperscript{1} Requires the input to not have any residual connections.  
    \noindent\textsuperscript{2} Software mapping needed to map FlexACC primitives in C.
    \noindent\textsuperscript{3} Manuscript description is ambiguous. 
    \noindent DSL = Domain Specific Language (PyTorch/TensorFlow, etc.)
    \end{threeparttable}
\end{table*}

\par
\begin{figure}
    \centering
    \includegraphics[width=\linewidth]{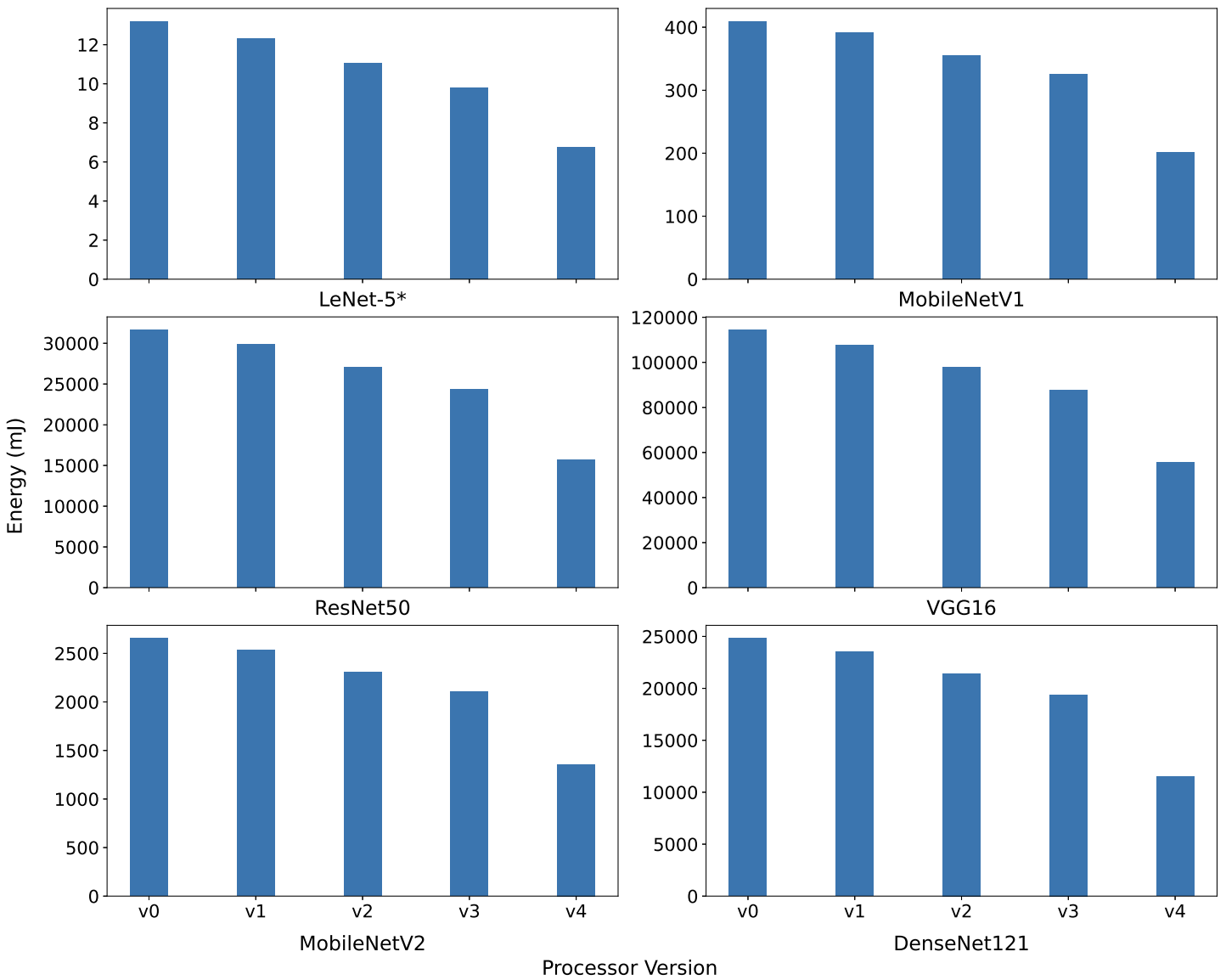}
    \caption{Energy/Inference on RISC-V variants}
    \label{energy_per_inf}
    \vspace{-4mm}
\end{figure}

 Energy efficiency is estimated for each model on each RISC-V variant as per (\ref{energy_eqn})
    \begin{equation}
    E = P \times \left( \frac{C}{f} \right)
    \label{energy_eqn}
\end{equation}

\noindent where E is the energy per inference, P is the estimated design power consumption, C is the number of clock cycles per inference and f is the processor clock frequency. In this design, the processor clock frequency is 100 MHz. While Synopsys trv cores target 80 MHz for prototyping by default, we evaluated 100 MHz to validate timing of processor version v4. This frequency was chosen as it meets timing without violations and requires no RTL modifications. The power utilisation is estimated by Xilinx Vivado based on the post-implementation design. Fig. \ref{energy_per_inf} shows how the extensions reduce the energy consumed per CNN inference over different models.

\section{Conclusion}
This paper introduced MARVEL, an automated end-to-end framework that generates custom RISC-V ISA extensions specifically optimized for convolutional neural networks (CNNs) targeting resource-constrained IoT devices. MARVEL bridges high-level Python-based AI models and low-level bare-metal C implementations, enabling efficient deployment without operating system dependencies. The proposed framework demonstrated up to 2× improvement in inference speed and energy efficiency with minimal hardware overhead (28.23\% area increase) compared to a baseline RISC-V core for popular DNN models. Future work includes refining power estimation accuracy, exploring additional RISC-V baselines, and extending support for diverse deep learning model classes with support for additional quantization levels, thereby enhancing edge AI hardware-software co-design.

\section*{Acknowledgments}
This material is based upon work supported by the National Science Foundation under Grant No.\ 2315851 and 2106634, the Department for the Economy, Northern Ireland under the US-Ireland R\&D Partnership Programme, grant agreement USI-226, the MicroElectronic Circuit Centre Ireland (MCCI-2020-07) and Research Ireland (22/US/3848).

\bibliographystyle{IEEEtran}
\bibliography{thebibliography}

\begin{thebibliography}{10}
\providecommand{\url}[1]{#1}
\csname url@samestyle\endcsname
\providecommand{\newblock}{\relax}
\providecommand{\bibinfo}[2]{#2}
\providecommand{\BIBentrySTDinterwordspacing}{\spaceskip=0pt\relax}
\providecommand{\BIBentryALTinterwordstretchfactor}{4}
\providecommand{\BIBentryALTinterwordspacing}{\spaceskip=\fontdimen2\font plus
\BIBentryALTinterwordstretchfactor\fontdimen3\font minus \fontdimen4\font\relax}
\providecommand{\BIBforeignlanguage}[2]{{%
\expandafter\ifx\csname l@#1\endcsname\relax
\typeout{** WARNING: IEEEtran.bst: No hyphenation pattern has been}%
\typeout{** loaded for the language `#1'. Using the pattern for}%
\typeout{** the default language instead.}%
\else
\language=\csname l@#1\endcsname
\fi
#2}}
\providecommand{\BIBdecl}{\relax}
\BIBdecl

\bibitem{SINGH202371}
R.~Singh and S.~S. Gill, ``Edge ai: A survey,'' \emph{Internet of Things and Cyber-Physical Systems}, vol.~3, pp. 71--92, Feb. 2023.

\bibitem{10488759}
E.~Guthmuller \emph{et~al.}, ``Xvpfloat: Risc-v isa extension for variable extended precision floating point computation,'' \emph{IEEE Transactions on Computers}, vol.~73, no.~7, pp. 1683--1697, 2024.

\bibitem{10.1145/3675018.3675029}
H.~Liu \emph{et~al.}, ``Convex: A risc-v instruction set extension scheme for accelerating convolution operations on mcus,'' in \emph{Proceedings of the 2024 8th International Conference on High Performance Compilation, Computing and Communications}, ser. HP3C '24.\hskip 1em plus 0.5em minus 0.4em\relax Association for Computing Machinery, 2024, p. 133–138.

\bibitem{10.1145/3587135.3592168}
R.~Sahita \emph{et~al.}, ``Cove: Towards confidential computing on risc-v platforms,'' in \emph{Proceedings of the 20th ACM International Conference on Computing Frontiers}, ser. CF '23.\hskip 1em plus 0.5em minus 0.4em\relax New York, NY, USA: Association for Computing Machinery, 2023, p. 315–321.

\bibitem{10.1145/3566097.3567872}
M.~Askarihemmat \emph{et~al.}, ``Barvinn: Arbitrary precision dnn accelerator controlled by a risc-v cpu,'' in \emph{Proceedings of the 28th Asia and South Pacific Design Automation Conference}, ser. ASPDAC '23.\hskip 1em plus 0.5em minus 0.4em\relax New York, NY, USA: Association for Computing Machinery, 2023, p. 483–489.

\bibitem{9658643}
Q.~Jiao, W.~Hu, F.~Liu, and Y.~Dong, ``Risc-vtf: Risc-v based extended instruction set for transformer,'' in \emph{2021 IEEE International Conference on Systems, Man, and Cybernetics (SMC)}, 2021, pp. 1565--1570.

\bibitem{eyeris}
Y.-H. Chen \emph{et~al.}, ``Eyeriss: An energy-efficient reconfigurable accelerator for deep convolutional neural networks,'' \emph{IEEE Journal of Solid-State Circuits}, vol.~52, no.~1, pp. 127--138, 2017.

\bibitem{tops}
D.~Kadetotad \emph{et~al.}, ``An 8.93 tops/w lstm recurrent neural network accelerator featuring hierarchical coarse-grain sparsity for on-device speech recognition,'' \emph{IEEE Journal of Solid-State Circuits}, vol.~55, no.~7, pp. 1877--1887, 2020.

\bibitem{7864441}
M.~Gautschi \emph{et~al.}, ``Near-threshold risc-v core with dsp extensions for scalable iot endpoint devices,'' \emph{IEEE Transactions on Very Large Scale Integration (VLSI) Systems}, vol.~25, no.~10, pp. 2700--2713, 2017.

\bibitem{xpulp}
A.~Garofalo, G.~Tagliavini, F.~Conti, D.~Rossi, and L.~Benini, ``Xpulpnn: Accelerating quantized neural networks on risc-v processors through isa extensions,'' in \emph{2020 Design, Automation and Test in Europe Conference and Exhibition 2020}, 2020, pp. 186--191.

\bibitem{cnnrisc}
X.~Yu \emph{et~al.}, ``Cnn specific isa extensions based on risc-v processors,'' in \emph{2022 5th International Conference on Circuits, Systems and Simulation (ICCSS)}, 2022, pp. 116--120.

\bibitem{10737828}
A.~Al-Qawlaq, M.~Ajay~Kumar, and D.~John, ``Kwt-tiny: Risc-v accelerated, embedded keyword spotting transformer,'' in \emph{2024 IEEE 37th International System-on-Chip Conference (SOCC)}, 2024, pp. 1--6.

\bibitem{Berg_2021}
A.~Berg \emph{et~al.}, ``Keyword transformer: A self-attention model for keyword spotting,'' in \emph{Proceedings of Interspeech 2021}, August 2021.

\bibitem{9516466}
E.-Y. Yang \emph{et~al.}, ``Flexacc: A programmable accelerator with application-specific isa for flexible deep neural network inference,'' in \emph{2021 IEEE 32nd International Conference on Application-specific Systems, Architectures and Processors (ASAP)}, 2021, pp. 266--273.

\bibitem{10.1145/3665283.3665342}
A.~Kumar~M, V.~Kumar, D.~John, and S.~Shanker, ``Implementation and analysis of custom instructions on risc-v for edge-ai applications,'' in \emph{Proceedings of the 14th International Symposium on Highly Efficient Accelerators and Reconfigurable Technologies}, ser. HEART '24.\hskip 1em plus 0.5em minus 0.4em\relax ACM, 2024, p. 126–129.

\bibitem{10909993}
G.~Korol and A.~C.~S. Beck, ``{ IoT–Edge Splitting With Pruned Early-Exit CNNs for Adaptive Inference },'' \emph{IEEE Transactions on Very Large Scale Integration (VLSI) Systems}, no.~01, pp. 1--0, Mar. 5555.

\bibitem{10463613}
H.~Sang \emph{et~al.}, ``An 2.31uj/inference ultra-low power always-on event-driven ai-iot soc with switchable nvsram compute-in-memory macro,'' \emph{IEEE Transactions on Circuits and Systems II: Express Briefs}, vol.~71, no.~5, pp. 2534--2538, 2024.

\bibitem{verma2022ai}
V.~Verma, ``Ai-risc: Scalable risc-v processor for iot edge ai applications,'' \emph{University of Virginia, PHD Thesis}, 2022.

\bibitem{asip_designer}
\BIBentryALTinterwordspacing
{Synopsys ASIP Designer}. [Online]. Available: \url{https://www.synopsys.com/dw/ipdir.php?ds=asip-designer}
\BIBentrySTDinterwordspacing

\bibitem{trv32p3}
\BIBentryALTinterwordspacing
{trv32p3 processor core}. [Online]. Available: \url{https://www.synopsys.com/designware-ip/processor-solutions/asips-tools/asip-models.html#trv}
\BIBentrySTDinterwordspacing

\bibitem{10.5555/3291168.3291211}
T.~Chen \emph{et~al.}, ``Tvm: an automated end-to-end optimizing compiler for deep learning,'' in \emph{Proceedings of the 13th USENIX Conference on Operating Systems Design and Implementation}, ser. OSDI'18.\hskip 1em plus 0.5em minus 0.4em\relax USA: USENIX Association, 2018, p. 579–594.

\bibitem{edgecortix}
\BIBentryALTinterwordspacing
{EDGECORTIX}, ``Mera compiler and software framework.'' [Online]. Available: \url{https://www.edgecortix.com/en/products/mera}
\BIBentrySTDinterwordspacing

\bibitem{iree}
\BIBentryALTinterwordspacing
``Intermediate representation execution envioonment.'' [Online]. Available: \url{https://github.com/iree-org/iree?tab=readme-ov-file}
\BIBentrySTDinterwordspacing

\bibitem{6755945}
J.~Krause, M.~Stark, J.~Deng, and L.~Fei-Fei, ``3d object representations for fine-grained categorization,'' in \emph{2013 IEEE International Conference on Computer Vision Workshops}, 2013, pp. 554--561.

\bibitem{cocodataset}
\BIBentryALTinterwordspacing
{COCO dataset}. [Online]. Available: \url{https://cocodataset.org/}
\BIBentrySTDinterwordspacing

\bibitem{tflite}
\BIBentryALTinterwordspacing
{LiteRT overview}. [Online]. Available: \url{https://ai.google.dev/edge/litert}
\BIBentrySTDinterwordspacing

\bibitem{LIANG2021370}
T.~Liang \emph{et~al.}, ``Pruning and quantization for deep neural network acceleration: A survey,'' \emph{Neurocomputing}, vol. 461, pp. 370--403, 2021.

\bibitem{nml-to-hdl_compiler}
\BIBentryALTinterwordspacing
{Synopsys Go HDL Generator Manual}. [Online]. Available: \url{https://spdocs.synopsys.com/dow_retrieve/latest/sg/asip_designer/go-manual.pdf}
\BIBentrySTDinterwordspacing

\bibitem{10257058}
Y.~Gao \emph{et~al.}, ``Liteair5: A system-level framework for the design and modeling of ai-extended risc-v cores,'' in \emph{2023 IEEE 36th International System-on-Chip Conference (SOCC)}, 2023, pp. 1--6.

\bibitem{10752449}
M.~Ali \emph{et~al.}, ``Rv-proviler: Evaluating risc-v isa for application-specific requirements,'' in \emph{2024 IEEE Nordic Circuits and Systems Conference (NorCAS)}, 2024, pp. 1--7.

\bibitem{VERMA2022100742}
V.~Verma, T.~{Tracy II}, and M.~R. Stan, ``Extrem-edge—extensions to risc-v for energy-efficient ml inference at the edge of iot,'' \emph{Sustainable Computing: Informatics and Systems}, vol.~35, p. 100742, 2022.

\bibitem{chess_compiler}
\BIBentryALTinterwordspacing
{Synopsys Chess Compiler Manual}. [Online]. Available: \url{https://spdocs.synopsys.com/dow_retrieve/latest/sg/asip_designer/chess_user-manual.pdf}
\BIBentrySTDinterwordspacing

\end{thebibliography}

\begin{IEEEbiography}
[{\includegraphics[width=1in,height=1.0in]{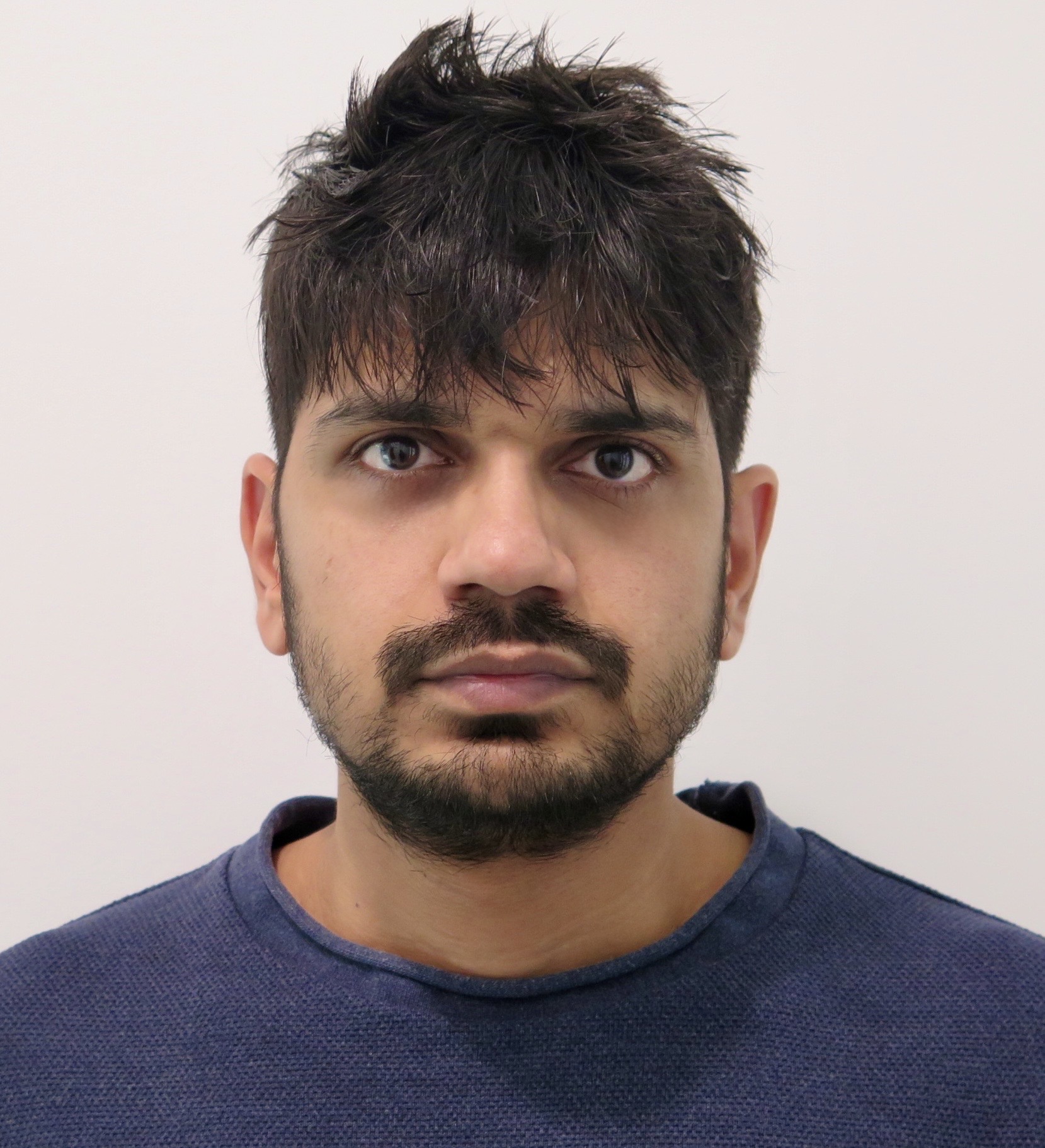}}] {Ajay Kumar M}\hspace{3pt}received the B.Tech degree in Electronics and Communication Engineering from National Institute of Technology Karnataka, India, in 2017. During 2017-2022, he worked with Texas Instruments India and Intel India. He is currently a PhD student with the School of Electrical and Electronic Engineering, University College Dublin, Ireland. His research interests include Embedded Systems, Edge-AI, Digital hardware design, FPGA, and DNN accelerators.
\end{IEEEbiography}

\begin{IEEEbiography}
[{\includegraphics[width=1in,height=1.0in]{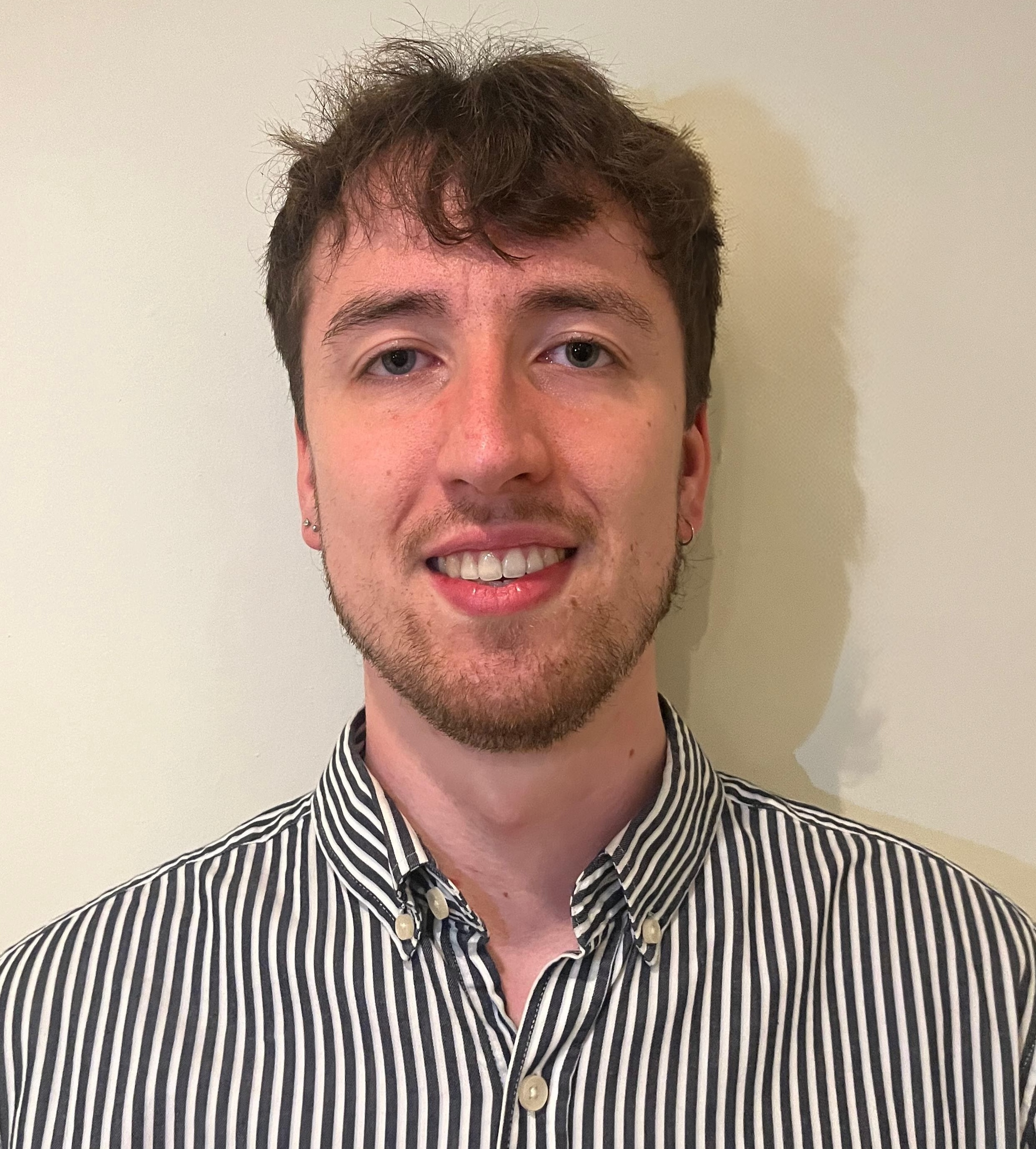}}] {Cian O'Mahoney}\hspace{3pt}received a B.Sc. degree in electronic and electrical engineering from University College Dublin, Dublin, Ireland, in 2023 and a M.E. in electronic and computer engineering from University College Dublin, Dublin, Ireland, in 2024. He is currently working as an FPGA Engineer in Susquehanna International Group LLP, Dublin, Ireland.
\end{IEEEbiography}

\begin{IEEEbiography}
[{\includegraphics[width=1in,height=1.0in]{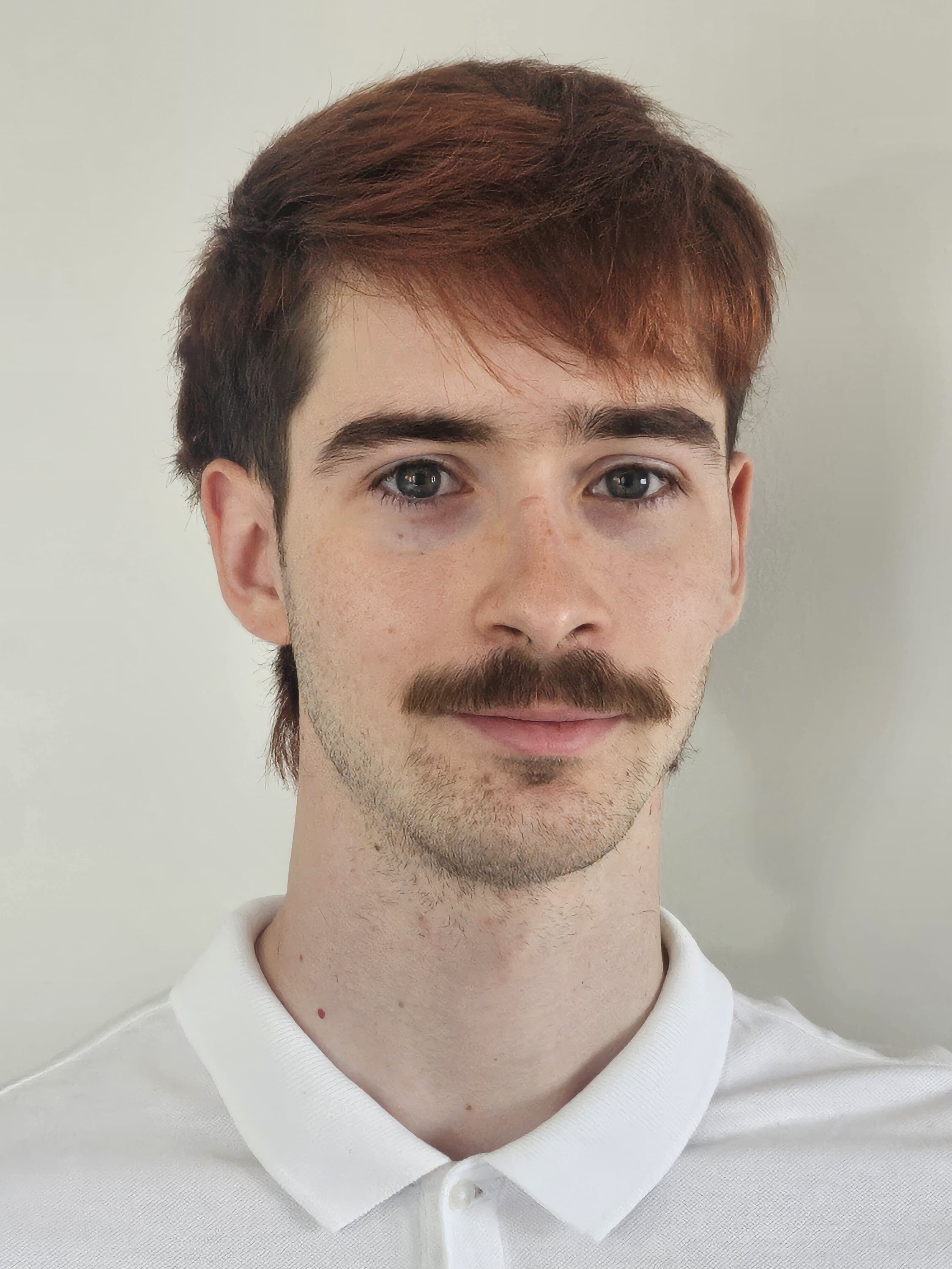}}]{Pedro Kreutz Werle}\hspace{3pt}received the B.Sc. and M.Eng. degrees in electronic and computer engineering from University College Dublin, Dublin, Ireland, in 2024 and 2025, respectively. In 2024/2025 he worked at the Intelligent Systems Lab at the University College Dublin as a Graduate Researcher and join as a Research Assistant. His research interests include embedded AI and energy-efficient architectures.

\end{IEEEbiography}

\begin{IEEEbiography}
[{\includegraphics[width=1in,height=1.0in]{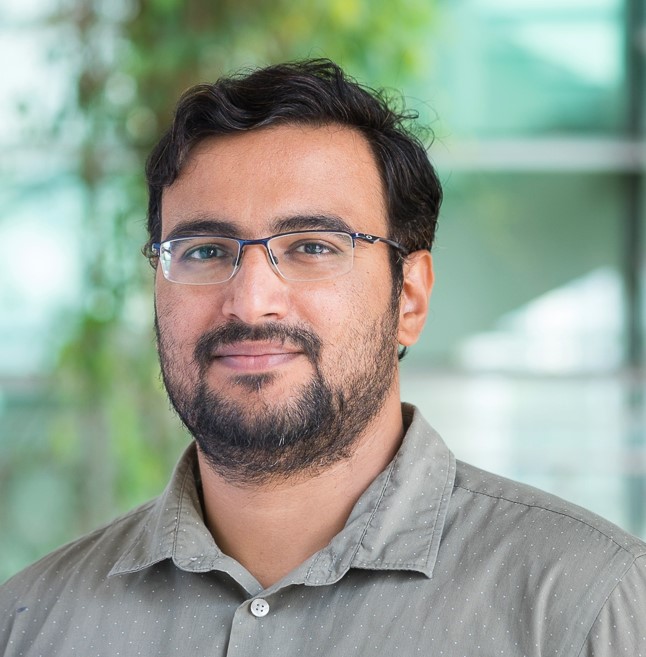}}]
{Shreejith Shanker}\hspace{3pt}is an Assistant Professor in Reconfigurable Computing at the School of Engineering, Trinity College Dublin, Ireland. He obtained his B.Tech. degree in Electronics and Communication Engineering from the University of Kerala, India and his Ph.D. degree in Computer Science and Engineering from Nanyang Technological University (NTU), Singapore, in 2006 and 2016, respectively.  
From 2006 to 2008, he worked as an FPGA design and development engineer. From 2008 to 2011, he
worked as a scientist at Digital Systems Group, Vikram Sarbhai Space Centre, Trivandrum, under the
Indian Space Research Organisation (ISRO). From 2015 to 2016, he was a Research Fellow at the School of Computer Science and Engineering, NTU, Singapore. From 2017 to 2018, he was a Teaching Fellow
at the School of Engineering, University of Warwick, UK. From 2018 to 2019, he was a Research Fellow at TUM CREATE Ltd, Singapore. Since 2019, he has been an Assistant Professor at the School of Engineering, Trinity College Dublin. 
He has also served as a member of the organising or technical committee for several IEEE/ACM conferences, such as FPL, DATE, FPT, ASAP, and HEART. He is also a reviewer for IEEE/ACM journals and conferences.
His research interests include reconfigurable and adaptive accelerators, distributed embedded systems, and in-network computing.
\end{IEEEbiography}

\begin{IEEEbiography}    
[{\includegraphics[width=1in,height=1.0in]{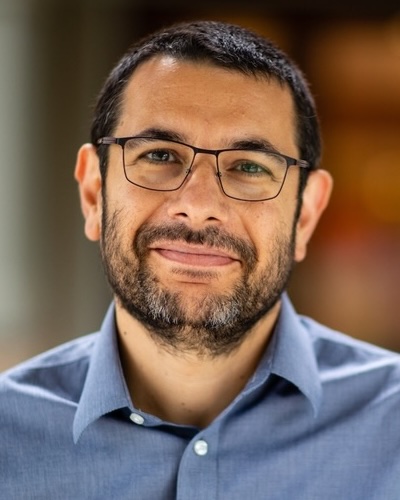}}]{Dimitrios~S. Nikolopoulos}\hspace{3pt}is the John W. Hancock Professor of Engineering at Virginia Tech, with appointments in the Department of Computer Science and the Department of Electrical and Computer Engineering. He received a Ph.D. in Computer Engineering from the University of Patras, Greece, in 2000. His research interests include high-performance computing, parallel and distributed systems, computer architecture, and systems software. He is an IEEE Fellow.
\end{IEEEbiography}

\begin{IEEEbiography}    
[{\includegraphics[width=1in,height=1.0in]{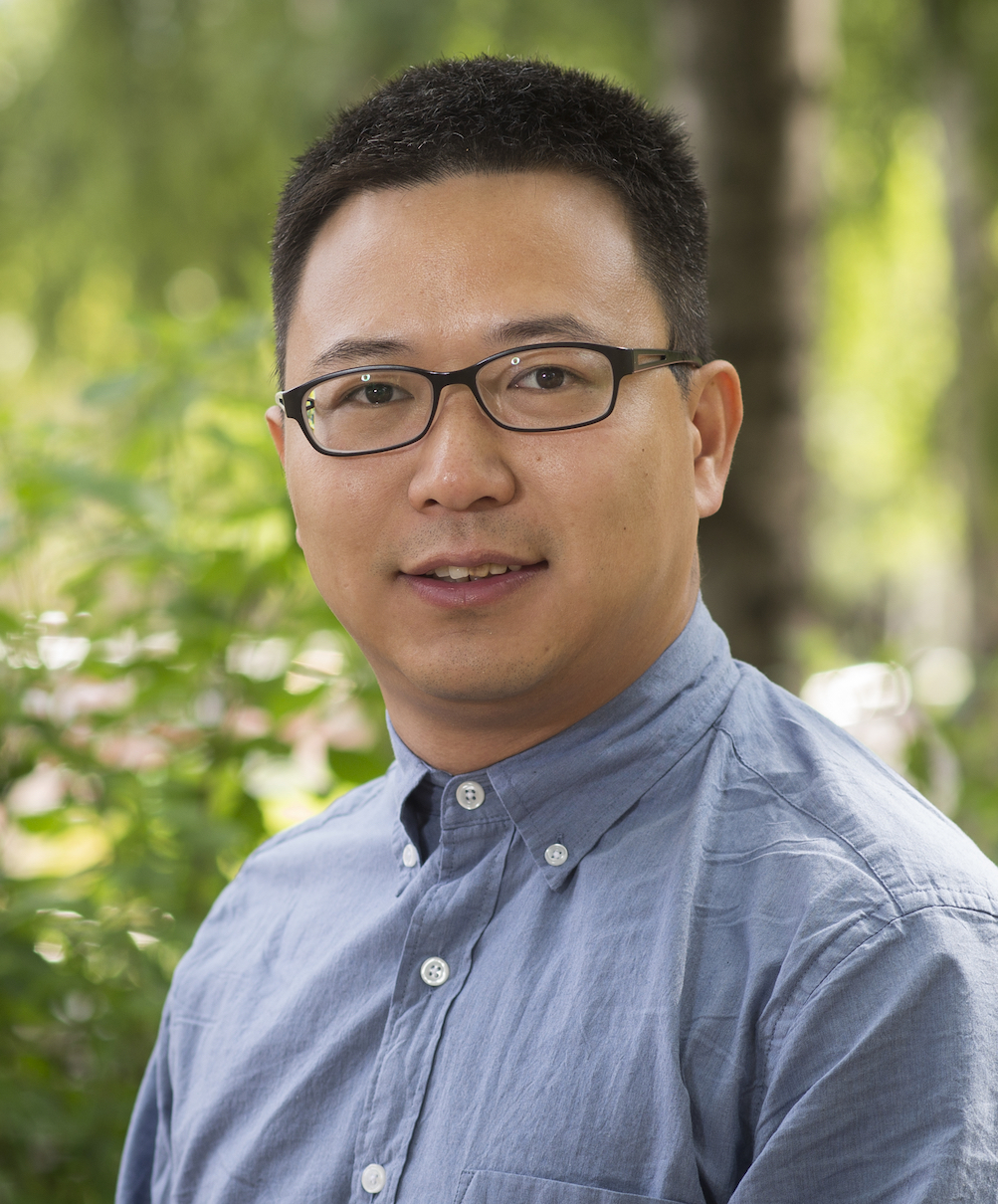}}]{Bo Ji}\hspace{3pt}received his Ph.D. degree in Electrical and Computer Engineering from The Ohio State University, Columbus, OH, USA, in 2012. Dr. Ji is an Associate Professor of Computer Science and a College of Engineering Faculty Fellow at Virginia Tech. Prior to joining Virginia Tech, he was an Associate Professor in the Department of Computer and Information Sciences at Temple University, where he was an Assistant Professor from July 2014 to June 2020. He was also a Senior Member of Technical Staff at AT\&T Labs, San Ramon, CA, from January 2013 to June 2014.
His research interests include the multidisciplinary intersections of Computing and Networking Systems, Artificial Intelligence and Machine Learning, Security and Privacy, and Extended Reality.
He has been the general co-chair of IEEE/IFIP WiOpt 2021 and the technical program co-chair of IEEE/IFIP WiOpt 2026, ACM MobiHoc 2023, and ITC 2021, and he has also served on the editorial boards of multiple IEEE and ACM journals (IEEE/ACM Transactions on Networking, ACM SIGMETRICS Performance Evaluation Review, IEEE Transactions on Network Science and Engineering, IEEE Internet of Things Journal, and IEEE Open Journal of the Communications Society). Dr. Ji is a senior member of the IEEE and the ACM and a member of the AAAS. He was a recipient of the National Science Foundation (NSF) CAREER Award in 2017, the NSF CISE Research Initiation Initiative Award in 2017, the IEEE INFOCOM 2019 Best Paper Award, the IEEE/IFIP WiOpt 2022 Best Student Paper Award, the IEEE TNSE Excellent Editor Award in 2021, 2022, and 2024, and the Dean's Faculty Fellow Award from the College of Engineering at Virginia Tech in 2023.
\end{IEEEbiography}

\begin{IEEEbiography}
[{\includegraphics[width=1in,height=1.0in]{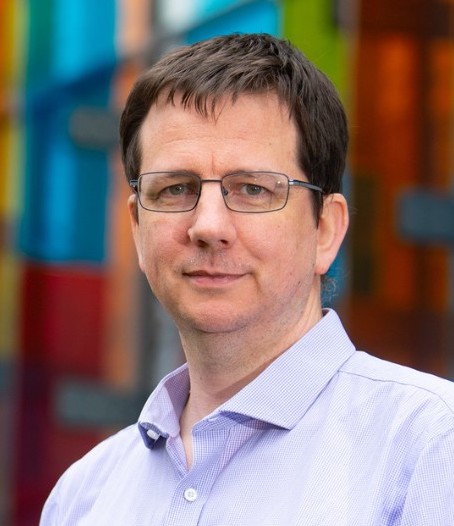}}]{Hans Vandierendonck}\hspace{3pt}received the Ph.D. and M.Sc. degrees from Ghent University, Ghent, Belgium, in 2004 and 1999, respectively. He is currently a Professor of high-performance and data-intensive computing with the School of Electronics, Electrical Engineering and Computer Science, and a Senior Member of ACM and IEEE.
His research aims to discover novel solutions to data-intensive parallel systems and algorithms.
\end{IEEEbiography}

\begin{IEEEbiography}
[{\includegraphics[width=1in,height=1.0in]{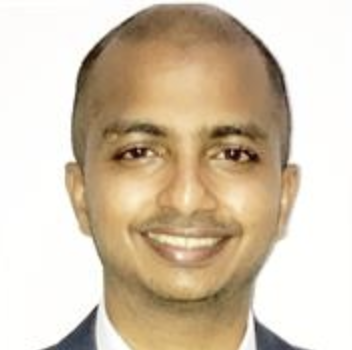}}]
{Deepu John}\hspace{3pt}is an Assistant Professor at the School of Electrical and Electronics Engineering, University College Dublin, Ireland. He received his Ph.D. in Electrical Engineering from the National University of Singapore in 2014. From 2014 to early 2017, he worked as a postdoctoral researcher at the Bio-Electronics Lab, National University of Singapore. Prior to this, he served as a senior engineer at Sanyo Semiconductors, Japan. He has received several awards, including the 2024 IEEE Transactions on Biomedical Circuits and Systems Best Paper Award, the Institution of Engineers Singapore Prestigious Engineering Achievement Award, Best Design Award at the Asian Solid-State Circuit Conference (2013), and IEEE Young Professionals, Region 10 individual award. He has served as an Associate Editor for IEEE Internet of Things Magazine and Guest Editor for IEEE Transactions on Circuits and Systems-I and IEEE Open Journal of Circuits and Systems. Currently, he serves as an Associate Editor for IEEE Transactions on Biomedical Circuits \& Systems, Wiley International Journal of Circuit Theory \& Applications and as a Senior Associate Editor for IEEE Transactions on Circuits \& Systems-II. His research interests include IoT/Wearable sensing, Biomedical Circuits \& Systems, and Edge Computing. He is a Senior Member of the IEEE.
 
\end{IEEEbiography}

\vfill\pagebreak

\end{document}